\documentclass[sigconf]{acmart}
\usepackage{algorithm}
\usepackage{algpseudocode}
\usepackage{enumitem}
\usepackage{colortbl}
\newcommand{\na}{\cellcolor{gray!20}—}
\usepackage{multirow}
\usepackage{rotating}
\usepackage{tcolorbox}
\tcbuselibrary{listings,breakable,skins}
\AtBeginDocument{%
  }

\setcopyright{acmlicensed}

\copyrightyear{2026}
\acmYear{2026}
\setcopyright{cc}
\setcctype{by}
\acmConference[UIST '26]{The 39th Annual ACM Symposium on User Interface Software and Technology}{November 02--05, 2026}{Detroit, MI, USA}
\acmBooktitle{The 39th Annual ACM Symposium on User Interface Software and Technology (UIST '26), November 02--05, 2026, Detroit, MI, USA}
\acmDOI{10.1145/3830398.3830671}
\acmISBN{979-8-4007-2856-3/2026/11}

\newcommand{\system}{\textsc{Compass}}
\newcommand\revise[1]{#1}

\begin{document}

\title{Compass: Continuously Aligning Social Media Feeds via In-Situ Reflections}

\author{Aadit Barua}
\authornote{Both authors contributed equally to this work.}
\email{aaditbarua@utexas.edu}
\affiliation{%
  \institution{The University of Texas at Austin}
  \city{Austin}
  \state{Texas}
  \country{United States}
}

\author{Leijie Wang}
\authornotemark[1]
\email{leijiew@cs.washington.edu}
\affiliation{%
  \institution{University of Washington}
  \city{Seattle}
  \state{Washington}
  \country{United States}
}

\author{Amy X. Zhang}
\email{axz@cs.uw.edu}
\affiliation{%
  \institution{University of Washington}
  \city{Seattle}
  \state{Washington}
  \country{United States}
}

\renewcommand{\shortauthors}{Barua et al.}

\begin{abstract}

Social media recommendation feeds often optimize for users’ immediate impulses rather than preferences they would hold after deeper reflection.
\revise{Some systems address this misalignment by incorporating users’ explicit preferences via a configuration page or in-feed controls instead of just behavioral signals.
However, users typically have evolving preferences, and their stated preferences and behavior naturally diverge, necessitating continuous reflection and feed realignment.
But existing strategies require the user to take initiative and are often effortful; as a result, in practice they are rarely invoked.}
We present \system{}, a system that aligns a user's feed with their reflective preferences by helping users reflect on and articulate their preferences given their behavior.
To enable continuous reflection during everyday browsing, \system{} surfaces in-situ reflections via lightweight notifications, while feed alignment is achieved by periodically simulating behavioral signals and directly manipulating feed content.
We embedded  \system{} within YouTube Shorts and compared it against a baseline without continuous support through a 10-day field study ($N=15$).
\revise{We found that \system{} promoted more reflective and purposeful feed consumption, iterative preference adjustment, and stronger feed alignment, without sacrificing the casual nature of feed browsing.}

\end{abstract}



\begin{teaserfigure}
    \centering
    \includegraphics[width=1.0\textwidth]{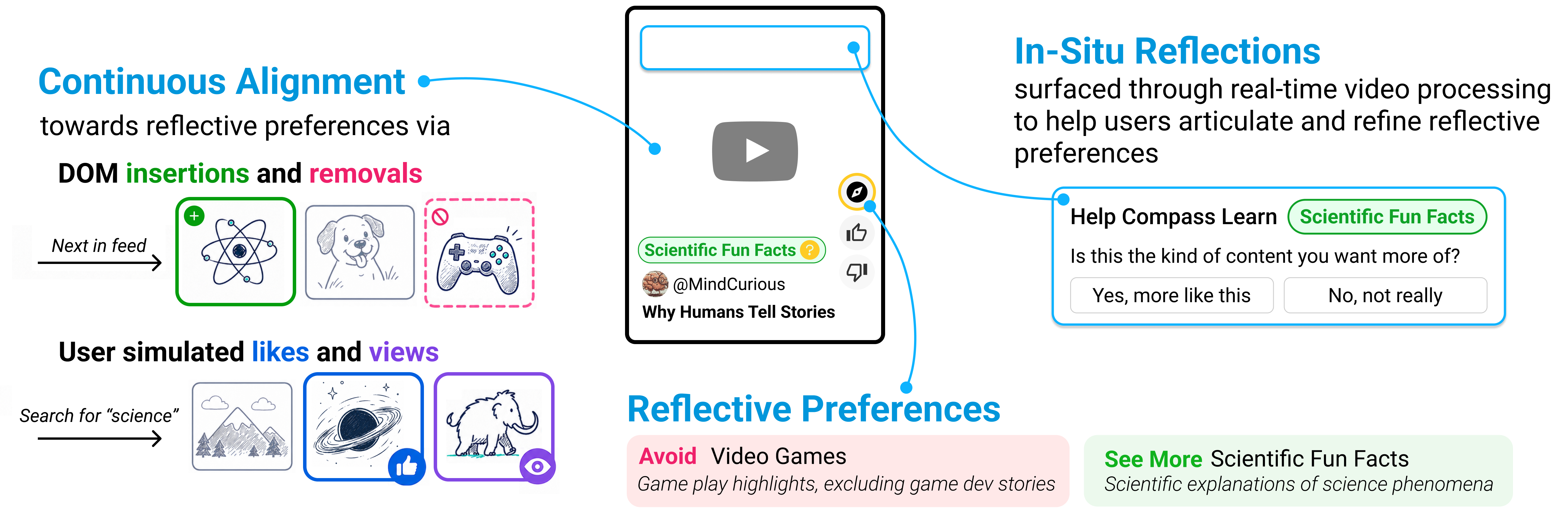}
    \caption{
        \revise{\system{} supports continuous alignment between social media recommendation feeds and users' reflective preferences.
        Rather than treating reflection as a one-time configuration, \system{} helps users articulate their reflective preferences through in-situ reflections during everyday browsing. \system{} then automatically aligns their feed through simulation of behavioral signals and direct DOM-based insertion and removal of content.}
        }
    \label{fig:teaser}
\end{teaserfigure}

\maketitle

\section{Introduction}


Recommendation algorithms on social media platforms are designed to maximize user engagement by learning from users’ behavioral signals (e.g., clicks, watch time, likes)~\cite{samuelson2024note, morewedge2023human, agan2023automating}. This leads to highly personalized feeds with little explicit effort from users~\cite{covington2016deep, vombatkere2024tiktok}. On short-form video platforms such as TikTok, YouTube Shorts, and Instagram Reels, fast-paced content and frictionless interaction make these systems especially responsive to users’ immediate reactions~\cite{zannettou2024analyzing}.
However, optimizing feeds around these signals can amplify attention-grabbing content in the moment, rather than content users actually want to spend their time on~\cite{brady2017emotion, brady2023algorithm}. As a result, users may find themselves engaging in ways they later regret~\cite{morewedge2023human}, while also contributing to broader issues such as addictive use~\cite{yang2025studying}, polarization~\cite{haroon2023auditing, brady2017emotion}, and misinformation~\cite{brady2023algorithm}. 

In response, researchers have explored ways to support users in shaping recommendation feeds toward more \textbf{reflective preferences}, or \textit{preferences that emerge through a process of reflection, where users consider their goals, values, and how they want their feeds to shape their time and attention}~\cite{ekstrand2016behaviorism, lukoff2021design,morewedge2023human, agan2023automating}.
One way to characterize the range of approaches is along a spectrum of how tightly they integrate with engagement-based algorithms.
At one end, some systems allow users to construct feeds from scratch by selecting sources or defining filtering rules~\cite{malki2025bonsai, choi2025designing}. Despite supporting expressive preference specification, these systems are often effortful to set up and lose the performance benefits of engagement-based algorithms that learn from large-scale behavioral data. 
Another line of work augments existing recommendation systems with additional configuration interfaces, allowing users to specify preferences through predefined categories or structured vocabularies, such as abstract values, political perspectives, or toxicity levels~\cite{kolluri2025alexandria, jia2024embedding, bhargava2019gobo}.
While these approaches retain the benefits of engagement-based recommendation, they require users to step outside the feed to configure their preferences and are often overlooked or forgotten over time~\cite{lukoff2018makes, hsu2020awareness, cunningham2024we}.

More importantly, these approaches largely treat shaping recommendation feeds as a one-time configuration task, overlooking the need for \textit{continuous} reflection and alignment.
In practice, users’ reflective preferences can evolve over time~\cite{carroll2022estimating}---meanwhile, behavioral signals are continuously steering engagement-based algorithms, causing feeds to drift away from any explicitly stated preferences~\cite{morewedge2023human, agan2023automating, kleinberg2024challenge}.
This creates two challenges.
First, prior systems assume that users will maintain a sustained awareness of their reflective preferences, notice moments of misalignment, and remember to act on them accordingly. 
But the highly engaging design of social media suppresses such moments; users rarely pause to articulate what feels misaligned and instead continue scrolling~\cite{baughan2022don, lukoff2018makes}.
Second, even when users do recognize a mismatch, prior systems provide limited support for translating these reflections into effective and sustained changes to the feed. Configuration-based approaches require users to manually encode their reflective preferences into structured settings and repeatedly update them over time~\cite{jia2024embedding, malki2025bonsai, kolluri2025alexandria}.
While platforms have introduced in-feed controls such as ``see more,'' or ``not interested,'' the effects of these actions are often difficult to observe or validate~\cite{vaccaro2018illusion, eslami2016first, devito2017algorithms}. Users must manually translate higher-level reflective preferences into repeated local actions~\cite{li2025beyond}, and as a result, these controls are significantly underused in practice~\cite{cunningham2024we, hsu2020awareness}.

We present \system{} (illustrated in Figure~\ref{fig:teaser}), a system that helps users continuously reflect on and align short-form video recommendation feeds with their evolving reflective preferences, implemented as a browser extension for YouTube Shorts.
First, \system{} surfaces opportunities for reflection \textit{in situ} to help users articulate their reflective preferences while scrolling through their feed. For instance, \system{} suggests new preferences by capturing moments that indicate potential misalignment during feed consumption and asks for in-feed annotations to incrementally refine its interpretation of these preferences.
These in-situ interactions are enabled by a real-time video processing pipeline to determine whether and how to intervene during videos as users browse. 
Second, \system{} continuously aligns users' recommendation feeds with their reflective preferences, as behavioral signals alone may otherwise steer recommendations away from what users want after reflection. To bridge the gap between users' reflective preferences and their actual behavior, \system{} translates expressed preferences into \textit{simulated} engagement behaviors to steer the underlying recommendation algorithm.  It also directly reshapes the presented feed through DOM manipulations that inject and remove content so that users immediately see the effects of their modifications regardless of the algorithm's responsiveness.

We evaluated \system{} through a between-subjects field study ($N = 15$) over 10 days comparing \system{} against a baseline where users must take initiative to specify their preferences and trigger the feed alignment process.
We found that in-situ reflections promoted more reflective social media use, with participants describing how they browsed more intentionally and purposefully. 
\revise{Critically, \system{} enabled significantly more frequent preference refinement (an average of $11.50$ vs.\ $0.14$ refinements per participant), while participants in the baseline condition were more likely to forget the system existed or lose track of their preferences over time.}
These additional refinements came without sacrificing the casual, discovery-driven nature of short-form video browsing (comparable SUS scores of $84.1$ vs.\ $83.2$).
\revise{\system{} also achieved stronger feed alignment than the baseline. \system{} feeds contained significantly more content that users wanted to see ($49.3\%$ vs.\ $20.3\%$), and \system{} users devoted a significantly larger share of their watchtime to it ($58.7\%$ vs.\ $21.0\%$), while maintaining comparable feed diversity.}
This was largely driven by \system{}'s proactive simulation of behavioral signals: baseline participants rarely triggered this simulation process themselves (averaging fewer than 2 manual triggers per participant), even though their feeds were less aligned with their preferences. We additionally report technical evaluations demonstrating that \system{}'s real-time video processing pipeline achieves strong relevance-matching performance suitable for in-feed intervention.

\revise{We conclude by discussing the broader implications of designing lightweight interventions for reflective feed usage, as well as how the techniques underlying \system{} may support other forms of user control over recommendation feeds.}

\section{Related Work}
\subsection{Aligning Recommendation Feeds with Stated Preferences}
Recommendation algorithms on social media platforms learn from users' behavioral signals (e.g., clicks, watch time, and likes) to infer what users want, a practice grounded in the economic notion of \textit{revealed preferences}~\cite{samuelson2024note, morewedge2023human, agan2023automating}. 
While feeds curated this way retain users longer than chronological ones~\cite{guess2023social}, revealed preferences can diverge from \textit{stated preferences}---what users explicitly express when asked~\cite{morewedge2023human, agan2023automating}. This is because behavioral signals can be subject to well-documented cognitive biases and are thus an imperfect proxy for genuine user interest~\cite{kahneman2011thinking, boonprakong2025hci}. As such, recommendation algorithms can prioritize emotionally charged and morally valenced content~\cite{brady2017emotion, brady2023algorithm}, contributing to issues ranging from societal harms such as polarization~\cite{haroon2023auditing, brady2017emotion} and misinformation~\cite{brady2023algorithm} to individual harms such as regret after use~\cite{morewedge2023human}.

Researchers and practitioners have tried varied approaches to incorporate stated preferences into recommendation systems. 
One line of systems support users to construct feeds entirely from stated preferences. 
Early work let users specify content sources, write filters, and select topics independently of platform algorithms~\cite{bhargava2019gobo}.
With the introduction of decentralized platforms such as Bluesky and Mastodon, there has been renewed interest in this paradigm~\cite{malki2025bonsai, liu2025understanding, choi2025designing, popowski2026social}. 
However, building a feed from scratch requires significant effort and loses the personalization benefits of engagement-based algorithms.
Not surprisingly, only a small fraction of users on platforms that support custom feeds ever builds one~\cite{quelle2025bluesky}.

In contrast, another line of work augments existing recommendation feeds with configuration interfaces. These systems typically post-hoc manipulate the feed through re-ranking, remixing, and filtering. These approaches vary in the expressiveness of their preference language. The most constrained rely on keyword or rule-based filtering~\cite{jhaver2023personalizing, choi2025designing}. More expressive approaches let users specify preferences through sliders or filters over predefined categories such as topics, abstract values, or partisanship~\cite{kolluri2025alexandria, jia2024embedding}, increasingly relying on LLM-based annotation to match feed content against these categories. More recently, LLMs have further offered the potential for users to directly express their nuanced stated preferences in natural language~\cite{carroll2025ctrl, tang2025interactive}. However, communicating preferences accurately remains a challenge, and some work has proposed using concrete content examples to help users articulate them more precisely~\cite{feng2024mapping, wang2025end}. Across all these approaches, a shared limitation remains: they rely on standalone configuration interfaces outside of the feed. Much like privacy settings~\cite{liu2011analyzing}, these interfaces are rarely discovered in the first place~\cite{hsu2020awareness}, and even users who are initially motivated to configure them tend to gradually forget to revisit and maintain them over time~\cite{cunningham2024we, lukoff2018makes}.

\subsection{The Challenge of Continuous Reflection and Alignment}

Despite these diverse approaches, we argue that they share two assumptions that undermine their ability to support users in the long run.
\textbf{First, they assume that users already have a clear idea of their stated preferences, as if they can simply be stated on demand rather than requiring ongoing support to surface.} But what users actually want from their feeds can be ambiguous, shaped by competing goals and values that have not yet been fully worked out~\cite{ekstrand2016behaviorism, lukoff2021design, popowski2026social}. They also evolve as users' personal goals and values change through everyday use~\cite{carroll2022estimating}. 
Compounding this, the fast-paced, highly engaging design of short-form video platforms suppresses the very moments that could prompt users to articulate or even become aware of their preferences~\cite{mildner2021ethical}. When something feels misaligned, users rarely pause to articulate why and instead continue scrolling~\cite{baughan2022don}. Even users who open these apps with a specific purpose in mind quickly lose sight of it as they engage in passive consumption~\cite{lukoff2018makes}. 
As such, we use the term \textbf{reflective preferences} to distinguish this from stated preferences: rather than something users can state on demand, these are preferences that emerge through an ongoing reflective process, where users continuously consider their goals, values, and how they want their feeds to shape their time and attention~\cite{ekstrand2016behaviorism, lukoff2021design, stray2021beyond}.

\textbf{Second, these prior approaches largely treat shaping recommendation feeds as a one-time configuration task rather than an ongoing alignment process.} Yet reflective preferences evolve over time, and behavioral signals continuously steer the recommendation algorithms away from users' stated preferences~\cite{morewedge2023human, agan2023automating, kleinberg2024challenge}. Both forces require continuous realignment but prior systems offer no lightweight path to do so. Users must start from scratch each time: rebuilding feeds from new sources or rules~\cite{malki2025bonsai, choi2025designing} or revisiting configuration interfaces~\cite{kolluri2025alexandria, jia2024embedding, carroll2025ctrl}. Worse, users may not even notice when their feeds have drifted, leaving misalignment unaddressed~\cite{eslami2015always, rader2015understanding}. This burden of continuously reflecting and realigning helps explain why, despite their availability, these systems see little sustained use in practice~\cite{cunningham2024we, lukoff2018makes}.

Indeed, platform-native in-feed controls represent the closest attempt to support such ongoing alignment, offering lightweight ways to collect stated preferences directly within the feed. Some collect explicit satisfaction signals through in-feed surveys~\cite{cunningham2025ranking}.
But they are typically too sparse and feedback too general to be useful for modeling individual preferences~\cite{goodrow2021recommendation, gupta2021feedback}. Others expose item-level controls such as ``show more,'' ``show less,'' or ``not interested'' for users to signal preferences in the moment~\cite{meta2022customize, li2025beyond}. 
But because these item-level controls are not coupled with higher-level preference articulation, they provide limited scaffolding for continuous reflection or even awareness of one's content viewing patterns.
Even when users have clear higher-level preferences in mind, they are left guessing as to whether the system can infer these from item-level input~\cite{vaccaro2018illusion, rader2018explanations}.
Much like users who develop \revise{folk theories (informal mental models) of feed algorithms}~\cite{eslami2016first, devito2017algorithms} and perform actions such as deliberately rewatching videos to signal interest~\cite{li2025beyond}, users must resort to repeatedly firing item-level controls, with unclear understanding of their impact~\cite{meta2022customize, li2025beyond}.
As a result, shaping recommendation feeds through these controls remains demanding~\cite{eslami2019user, vera2025they} and significantly underused in practice~\cite{cunningham2024we, hsu2020awareness}.

\subsection{In-Situ Support to Promote Reflection}
A parallel body of work has explored supporting reflection and behavior change in habitual technology use, spanning smartphone use, fitness, and health~\cite{kim2016timeaware, kim2019goalkeeper, consolvo2008flowers, choe2015sleeptight}. We draw two design insights from this work. First, reflection is most effective when it occurs in the context of the behavior it targets. Prior work has designed ambient displays and lock-screen widgets that surface relevant information during use, rather than requiring users to seek out a separate interface~\cite{kim2016timeaware, choe2015sleeptight}. Second, such in-context interventions must strike the right balance between engaging users in reflection and not overwhelming them. Researchers have shown that automating the collection of behavioral data eliminates the reactive benefit of tracking, despite being accurate~\cite{nelson1981theoretical, choe2015sleeptight}, while overly intrusive interventions cause frustration and resistance~\cite{kim2019goalkeeper, ruiz2024design}. 
\revise{For short-form video platforms, recent work has explored placing interactive prompts toward the end of videos to spark curiosity and guide users into related content for exploration and incidental learning~\cite{tan2025curious}. However, it remains a central design challenge identifying when and how to surface timely and lightweight reflection over reflective preferences beyond curiosity amid fast-paced consumption.}

\begin{figure*}
    \centering
    \includegraphics[width=0.98\textwidth]{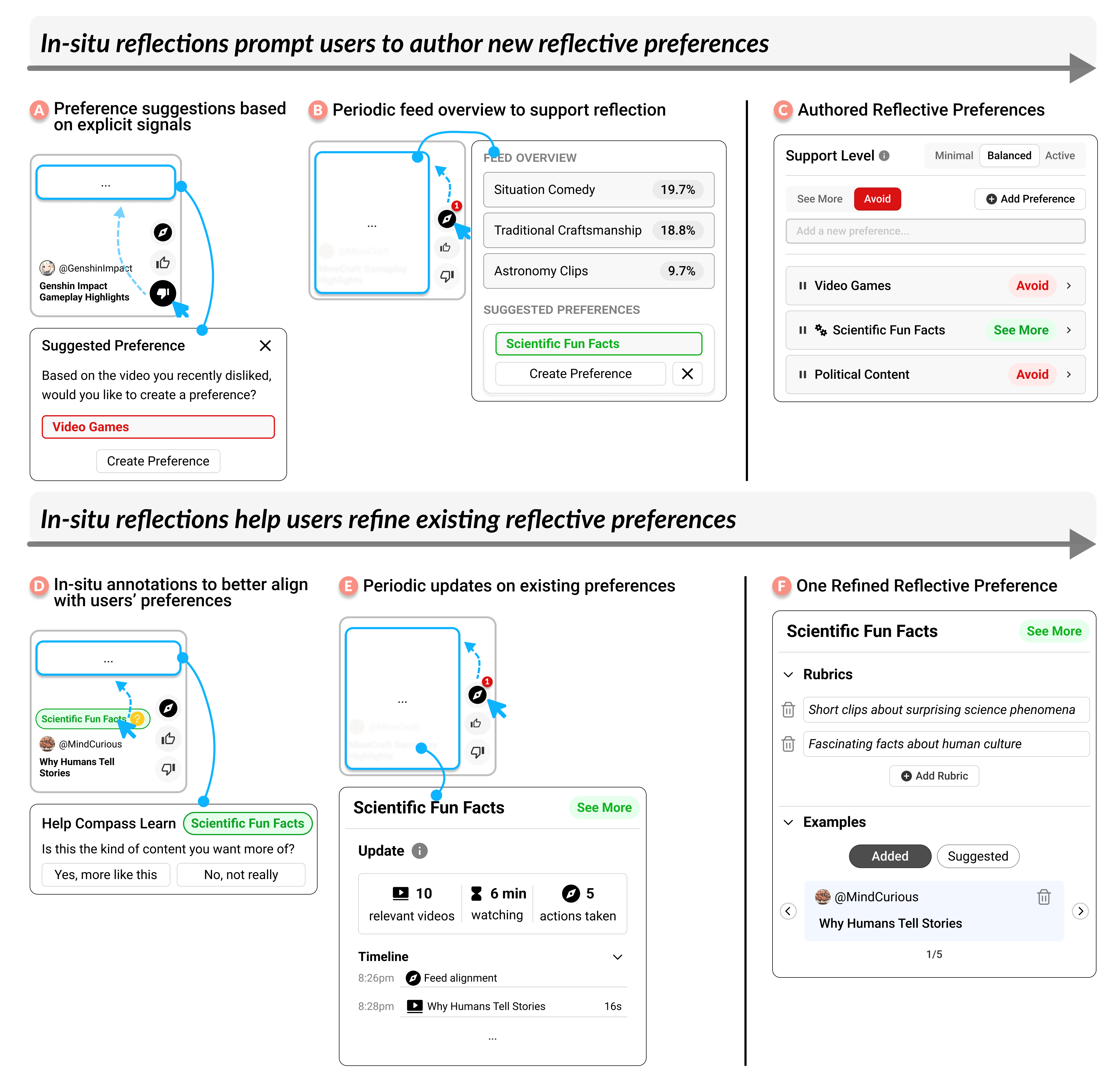}
    \caption{
    \textbf{\textsc{\system{}} embeds lightweight interactions directly into the recommendation feed to support users to author new reflective preferences and refine them more accurately over time, accumulating into a central preferences page for ongoing management.} 
    \textbf{(A-C) In-situ reflections prompt users to author new reflective preferences.}
    {\normalfont (A) When users express explicit signals during browsing (e.g., disliking a video), \system{} suggests new preferences that can be directly created in situ. (B) Periodic feed overviews summarize recent consumption patterns and surface suggested preferences based on recurring themes. (C) Newly created preferences are consolidated in the central preferences page, where users can manage them and adjust system support levels.}\\
    \textbf{(D–F) In-situ reflections support iterative refinement of existing preferences.}
    {\normalfont (D) \system{} collects in-situ annotations by asking users to label videos as relevant or irrelevant to an existing preference so that it can better interpret that preference over time. (E) The system provides periodic summaries of how existing preferences have shaped the feed, prompting users to reflect on whether those preferences still capture what they want and to update them when needed. (F) These interactions refine preference representations over time, including updated rubrics and examples maintained in the preference detail view.}
    }
    \label{insitu-interactions}
\end{figure*}

\section{Compass}


Building on the challenges and prior insights identified in the related work, we introduce three design goals:
\begin{itemize}[leftmargin=*]
\item
\textbf{DG1: Support expression of and reflection over preferences.} Rather than relying solely on behavioral signals such as watch time, users should be encouraged to explicitly articulate their preferences and thus reflect on whether their recommendations actually align with what they want to see.
\item
\textbf{DG2: Continuously align feeds with reflective preferences while preserving recommendation quality.} 
Users' feeds should be meaningfully shaped by their reflective preferences, and any changes should be clearly conveyed to the user. Meanwhile, any system should preserve the personalization benefits that make modern recommendation algorithms popular in the first place.
\item
\textbf{DG3: Preserve the low-effort, casual nature of everyday social media use.} 
Social media is typically used during breaks, commutes, or moments of downtime, making it an inherently low-effort activity. The system should fit naturally into this context, avoiding lengthy configuration tasks or overwhelming information that feel at odds with people's usage patterns.
\end{itemize}

To meet these design goals, we introduce \system{}, a system that helps users continuously reflect on their evolving preferences and align their recommendation feeds accordingly. We instantiated our system as a browser extension over YouTube Shorts, which is one of the largest short-form video platforms that averages 200 billion daily views~\cite{youtube_press_2026}. 
During implementation, we iteratively refined \system{} through a group pilot with lab members and individual pilots with four short-form video users.

\subsection{System Walkthrough}
To illustrate how \system{} works, consider Alice, a casual social media viewer of short-form videos who wants to see more scientific fun facts and also reduce her tendency to watch video game content that she finds overly engaging.

\subsubsection{\textbf{Support Lightweight Communication of Reflective Preferences}}
In \system{}, users create a preference by entering a title and specifying a direction (\textit{See More} or \textit{Avoid}).
As illustrated in Figure~\ref{insitu-interactions}c, Alice creates a \textit{See More} preference for scientific fun facts. 
After creation, she sees an initial system-generated rubric: a set of bullet points, each describing a category of relevant content (e.g., \textit{short clips about surprising science phenomena}).
This initial rubric may not fully capture Alice's nuanced preferences~\cite{wang2025end, kumar2023watch}.
For instance, she may realize her interest in fun facts extends beyond the natural sciences into humanities. 
Rather than requiring Alice to specify these nuances upfront, \system{} distributes this effort over time through requests for in-situ annotations during relevant videos. These interactions fit naturally into how casual social media users consume recommendation feeds by scanning and informally evaluating content as they browse~\cite{shen2021everyday}.

As shown in Figure~\ref{insitu-interactions}d, \system{} surfaces interactions during Alice's scrolling on a video about human culture where its match to her preference is uncertain,
asking whether she wants see more of such videos. With her response, \system{} then expands the rubric of her preference to also include fascinating facts about human culture (Figure~\ref{insitu-interactions}f).
It also presents less prominent indicators alongside videos that match the preference with high certainty, which Alice can expand to provide feedback if she thinks the match is incorrect.
These lightweight annotations allow \system{} to incrementally align with Alice's reflective preferences, without requiring Alice to stay vigilant about mismatches or proactively revise her preference.

At the same time, Alice may not know what preferences to specify upfront.
Prior work suggests moments of reflection often arise during feed consumption, when users express stated preferences through behavioral signals~\cite{li2025beyond}. 
\system{} builds on this by monitoring these signals to suggest new preferences.
For instance, in Figure~\ref{insitu-interactions}a, when Alice watches several video game clips and then dislikes one after realizing she has been drawn into them, the system surfaces an in-feed suggestion to create an \textit{Avoid} preference for \textit{Video Games}. Alice can then add it directly, expanding her preferences without leaving the feed.

\subsubsection{\textbf{Support Reflection at Varying Degrees of Prominence}}
While the in-situ interactions described above are tightly coupled to immediate browsing moments, \system{} also provides dedicated interactions for reflecting over broader patterns over time.
In Figure~\ref{insitu-interactions}e, \system{} periodically reminds Alice about updates to her reflective preferences. She can click the reminder to open an in-feed view summarizing how the system has acted on them. For example, her \textit{See More} preference about scientific fun facts has surfaced 10 matching videos in the past 48 hours, with a notably high share of humanities-related clips. Reviewing these summaries, Alice realizes she also enjoys history content and adds ``History Facts'' as another bullet point, broadening the preference scope to better reflect her evolving interests.
Additionally, \system{} periodically surfaces a summary of Alice's overall feed consumption, highlighting patterns across different content categories she has been watching, such as \textit{situation comedy, traditional craftsman videos, and astronomy clips} (Figure~\ref{insitu-interactions}b). Based on these patterns, \system{} proactively suggests new preferences that may better reflect her underlying interests. 

At a higher level, these in-feed interactions, spanning from video-level annotations to broader pattern summaries, not only help align recommendations with Alice's reflective preferences, but also act as lightweight reflection nudges that keep her aware of them.
These interactions are intentionally designed to vary in prominence, balancing awareness of reflective preferences with uninterrupted browsing. She can also manually adjust the overall frequency of these interactions to match her preferred level of engagement (Figure~\ref{insitu-interactions}c).


\subsubsection{\textbf{Automatically Align Recommendation Feeds with Reflective Preferences}}
\begin{figure}
    \centering
    \includegraphics[width=\columnwidth]{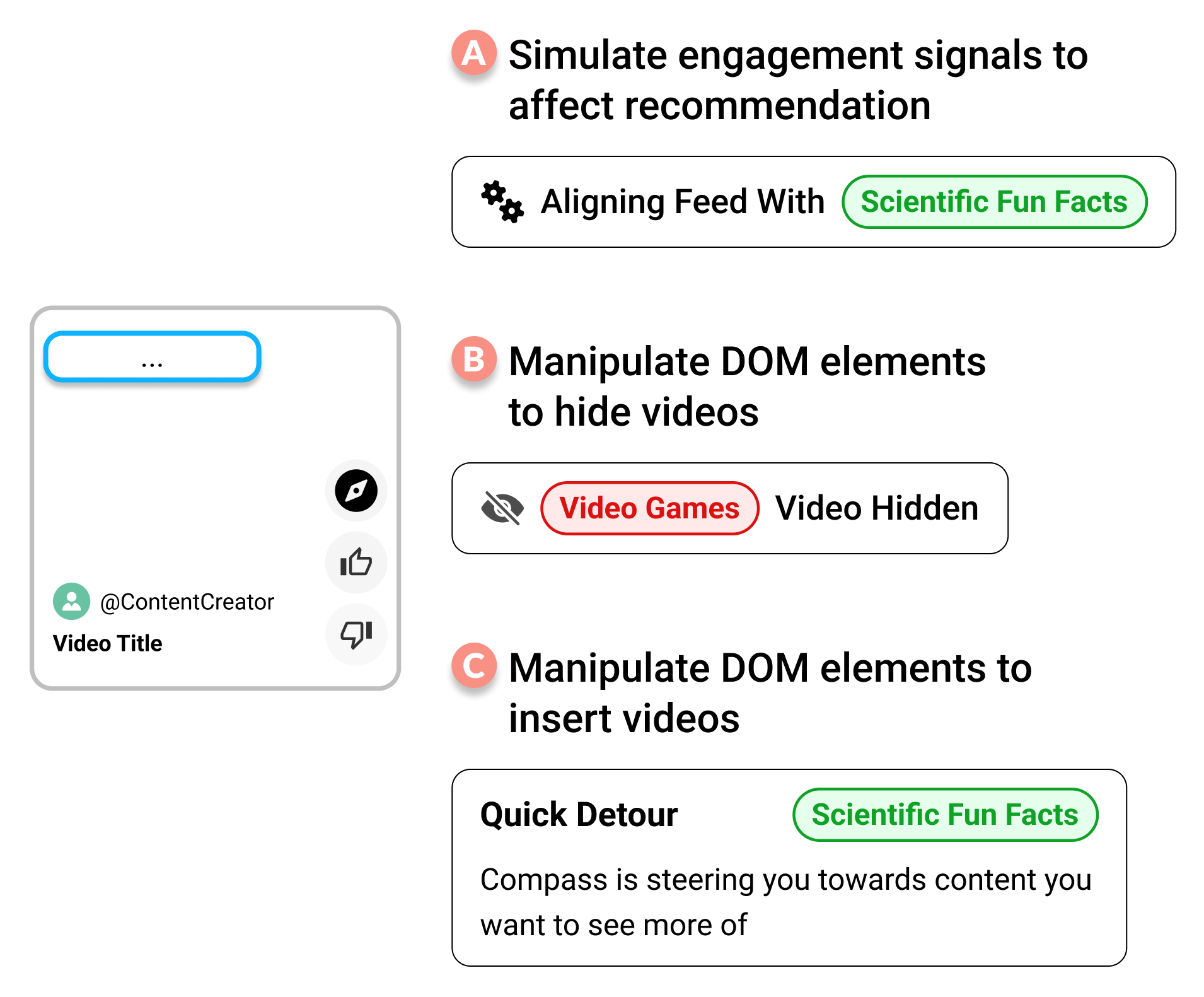}
    \caption{
        \textbf{Whenever it detects mismatches between the feed and users' reflective preferences, \system{} automatically realigns the feed, reducing the need for users to repeatedly monitor or manually correct recommendation drift. \system{} also displays ephemeral notifications to make these actions transparent.}
        \textbf{(A) Simulate engagement signals to steer recommendation.} {\normalfont \system{} simulates engagement behaviors in the background to influence future recommendations using the behavioral signals the recommender already relies on.}
        \textbf{(B) Manipulate DOM elements to hide videos.} {\normalfont For \textit{Avoid} preferences, \system{} removes matching videos from the feed before they are shown to the user. }
        \textbf{(C) Manipulate DOM elements to insert videos.} {\normalfont When recommendation steering alone is insufficient, \system{} directly inserts videos relevant to a \textit{See More} preference into the feed to provide immediately aligned content.}
    }
    \label{feed-alignment}
\end{figure}

As Alice scrolls through videos, \system{} continuously aligns her feed with her reflective preferences, since her behavioral signals can gradually steer recommendations away from them. This spares her from manually monitoring such drift and repeatedly intervening. Whenever \system{} aligns the feed, it presents temporary notifications to make its actions transparent to Alice (Figure~\ref{feed-alignment}). When the system detects that not enough clips about scientific fun facts have been recommended in her recent feed, it runs background simulations to steer the recommendation algorithm toward surfacing more such content. If it takes a while for the underlying algorithm to respond to such simulations, \system{} also injects relevant videos directly into the feed to ensure immediate exposure. For \textit{Avoid} preferences, such as video game content, matching content is filtered out before it reaches her.

\subsection{System Architecture}

To support this workflow, \system{} processes videos in real time, maintains evolving preference representations, and aligns recommendation feeds through simulation and direct manipulation.

\subsubsection{\textbf{Real-Time Video Processing Pipeline}}

A challenge for real-time feed intervention is processing videos quickly enough to act before users encounter them, a constraint that is particularly acute in short-form feeds where each item is consumed within only a few seconds. One observation that enables our system is that short-form video platforms typically preload a queue of upcoming videos rather than fetching them strictly one by one. We chose YouTube Shorts because its preloading API returns responses with interpretable structure: the \system{} extension intercepts these calls~\cite{piccardi2024reranking} and extracts video identifiers in advance, giving the backend time to process them before the user views them. 

Building on top of this, our video processing pipeline is designed to balance representation accuracy and processing latency, enabling feed interventions in real-time.
While more comprehensive approaches (e.g., downloading the full video to transcribe its audio and analyze all keyframes) could provide more accurate representations, they are too slow for such settings. Instead, we adopt a more lightweight representation. The pipeline first extracts the video's metadata (title, description, channel, tags, etc.) using a video downloading library.\footnote{yt-dlp: \url{https://github.com/yt-dlp/yt-dlp}} In parallel, it downloads three auto-generated thumbnails from YouTube that capture different moments in the video. We then pass both into a vision-language model to synthesize these complementary modalities into a unified semantic description. The generated description is then embedded.

\begin{figure*}
    \centering
    \includegraphics[width=0.75\textwidth]{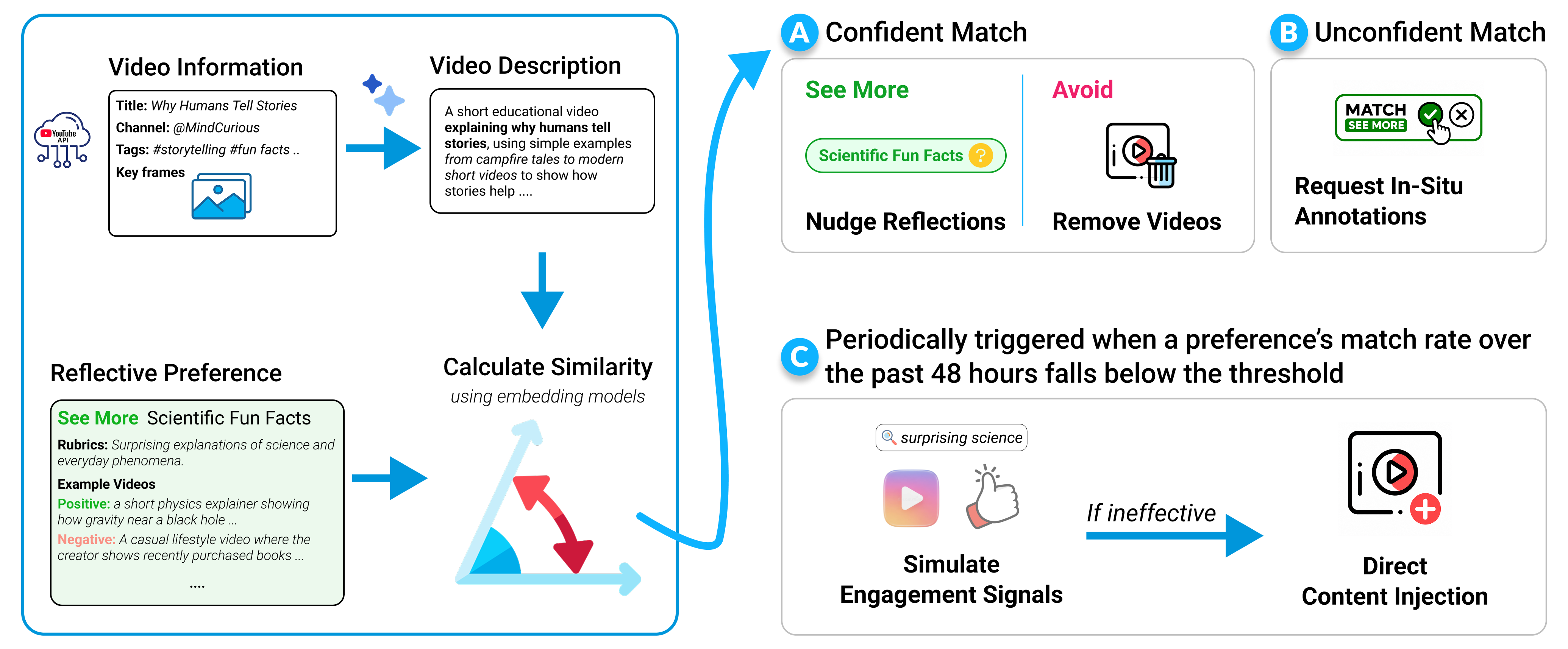}
    \caption{
    \revise{\textbf{\system{} semantically represents upcoming videos and users' reflective preferences, then compares these representations to support feed reflection and alignment.}
    {\normalfont
    For each upcoming video, \system{} retrieves its metadata and keyframes to generate a semantic description and then compares it against preference representations built from textual descriptions and example videos.
    The resulting similarity enables \system{} to perform appropriate system actions based on heuristic rules.
    For confident matches, it surfaces reflection nudges for ``See More'' preferences or removes videos matching ``Avoid'' ones; for unconfident matches, it requests in-situ user annotations.
    \system{} also periodically evaluates each preference's match rate over a sliding window.
    When the rate falls below a threshold, it first simulates engagement signals to steer the recommender and, if this proves ineffective, directly injects relevant videos into the feed.}
    }}
    \label{system-diagram}
\end{figure*}

\subsubsection{\textbf{Interpretable and Evolving Preference Representations}}
Each preference in \system{} is represented as a rubric of bullet points outlining categories of relevant content.
Rather than manually articulating these categories~\cite{wang2025end}, users can instead add positive or negative video examples through in-feed interactions, which trigger automatic updates to the rubric. We then embed each preference to enable direct comparison with video embeddings.

\system{} matches videos to preferences by first computing the similarity between their embeddings and then applying a threshold to determine whether a video matches a preference. This enables the system to surface in-feed interactions at appropriate moments. However, this matching assumes that the preference representation is well calibrated with the user's intent, which may not always hold. We therefore treat uncertain matches as opportunities to solicit user feedback through in-feed annotations. Drawing from uncertainty sampling in active learning~\cite{lewis1995sequential, seung1992query}, we flag videos with similarity scores near the threshold as uncertain, with the threshold set more strictly for preferences with fewer user-provided examples, reflecting lower confidence in those representations.


\subsubsection{\textbf{Aligning Recommendation Feeds via Simulation and Direct Manipulation}}
To align the feed with users' reflective preferences, \system{} operates on top of existing recommendation algorithms rather than replacing them with a new algorithm. This design leverages existing recommenders' ability to learn from behavioral signals while compensating for the gap between behavioral signals and reflective preferences~\cite{samuelson2024note, morewedge2023human, agan2023automating}. 
Prior work shows that users repeatedly like, dislike, or search for content in an attempt to bridge this gap~\cite{li2025beyond}. However, this process is not always reliable and requires continuous effort to monitor outcomes.

\system{} follows the same intuition but automates this translation process. When a \textit{See More} preference becomes underrepresented, the system simulates engagement behaviors to steer the recommendation system towards surfacing more matching videos. \revise{This happens when fewer than 6\% of videos in the user's feed over a sliding 48-hour window match the preference, a threshold determined through pilot experiments.}
Specifically, \system{} generates search queries for this preference, retrieves candidate videos, and simulates engagement such as watching and liking them in a background tab.
The system also monitors the resulting feed and continues simulation if the effect is limited.

However, recommendation algorithms do not always respond reliably or immediately to such signals. 
For example, prior work suggests that negative feedback on YouTube has limited impact~\cite{ricks2022youtube}.
Even when signals are effective, their effects may only manifest gradually, whereas users expect to see their preferences reflected in the feed immediately.
To address this, \system{} also applies direct manipulations to the feed when recommendation signals alone are insufficient. 
\revise{These manipulations can support alignment, but their primary value is giving users an immediate sense of feed alignment rather than waiting for simulation to take effect.}
For \textit{See More} preferences, it retrieves a pool of relevant videos based on the preference description. It then injects one unwatched video from the pool into the feed by updating the URL upon a user swipe, ensuring a smooth transition without disrupting browsing flow.
For \textit{Avoid} preferences, \system{} removes matching videos directly from the feed via DOM manipulation. Specifically, it locates any matching video in the DOM using its identifier and deletes it before it becomes visible to the user.
Together, these mechanisms allow \system{} to both steer and override recommendations when necessary, maintaining alignment with users' reflective preferences.

\subsection{Implementation Details}
\system{} is implemented as a client-server system consisting of a Chrome browser extension and a Python backend. The extension is built using Manifest V3 and vanilla JavaScript, with content scripts injected into YouTube Shorts pages and a service worker coordinating communication between the extension and backend. The backend uses FastAPI and SQLite, containerized with Docker to simplify installation. All model calls use the OpenAI API: \texttt{text embedding-3-large} for text embeddings, \texttt{GPT-4.1-nano} for low-latency video description generation, and \texttt{GPT-5.2} for all other text generation tasks. The full set of prompts is provided in Appendix~\ref{app:prompts}.

\section{Field Study}

We conducted a between-subjects field study comparing \system{} against a baseline with manual preference specification and alignment triggering out of the feed, isolating the effect of \system{}'s ongoing and in-situ support for reflection and alignment. We explore the following research questions:

\begin{itemize}
    \item[\textbf{R1:}] Does \system{} support continuous reflection on and 
    refinement of preferences during everyday social media use?
    \item[\textbf{R2:}] Does \system{} support continuous alignment of recommendation 
    feeds with users' preferences, while preserving the benefits of the underlying 
    recommendation algorithm?
\end{itemize}

\subsection{Methods}

\subsubsection{\textbf{Baseline System}} The baseline represents existing systems where users manually specify reflective preferences and apply them to the feed, without in-situ support for continuous reflection or alignment~\cite{kolluri2025alexandria, jia2024embedding, malki2025bonsai}.
We implement this baseline by stripping away these ongoing supports from \system{}, requiring users to proactively create and manage their preferences.
Upon creation, the system performs a one-time simulation to align the feeds with their \textit{See More} preferences, and also deploys filters that hide any content matching \textit{Avoid} preferences. As such, participants who have evolved \textit{See More} preferences or experience feed drifts need to manually re-trigger the alignment process via a dedicated button on the preference page.


\begin{figure*}
    \centering
    \includegraphics[width=\textwidth]{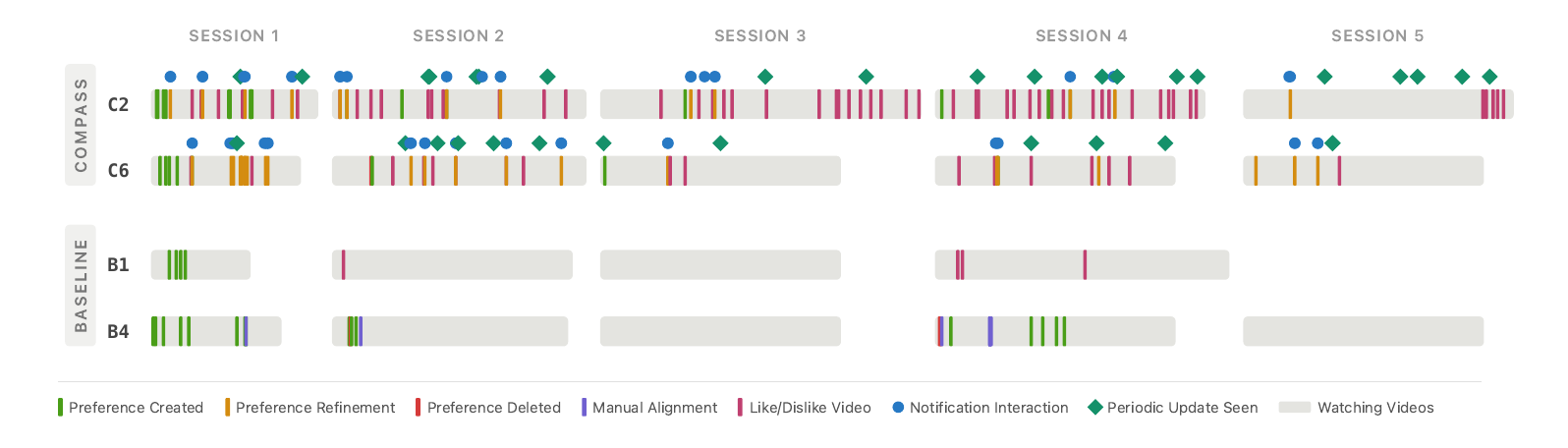}
    \caption{%
      \revise{\textbf{Session activity timelines for two \system{} participants (C2, C6) and two Baseline participants (B1, B4) across up to five study sessions.}
      {\normalfont Each gray bar represents the time a participant spent watching videos during a session. Colored marks on or above the bar indicate logged interaction \emph{events}.}
      {\normalfont Specifically, \emph{Notification interaction} refers to participants' proactive engagement with an in-situ reflection as opposed to passive dismissal.}
      {\normalfont Full session activity timelines for all participants are provided in Appendix Figure~\ref{usage-timeline}.\strut}}
  }
    \label{usage-timeline-focused}
\end{figure*}

\subsubsection{\textbf{Study Procedure}}
\revise{We recruited 15 social media users via personal connections and word of mouth, 
assigning 8 (C1--8) to \system{} and 7 (B1--7) to the baseline condition 
using alternating assignment.} We mainly targeted people age 18--24 as they are heavier users of short-form video platforms over older demographics~\cite{pew2025socialmedia}.
Each participant was compensated \$80 
(see Appendix Table~\ref{tab:participants} for further participant information). During the one-hour onboarding session, we first conducted a brief interview to collect demographic information and understand their typical social media use. We then assisted participants in installing their assigned system on their personal computer and introduced its features.

Participants were encouraged to engage in up to five sessions of system use over the course of 10 days, with each session lasting 30 minutes.
To approximate real-world use, we allowed sessions to be completed in one sitting or split across multiple sittings.
To ensure that the system was functioning properly, we asked participants to finish the first session independently during the onboarding session, so that we were available to address any setup issues.
To isolate the effect of their assigned system on YouTube's recommendation algorithm, participants were asked not to use YouTube or YouTube Shorts outside of study sessions. If they wished to do so, they were instructed to use a different YouTube account or incognito mode. After each session, participants filled out a form to report their experience. 
After completing the field study, participants attended a 45-minute session that included a SUS questionnaire~\cite{brooke1996sus}, a semi-structured post-study interview (see Appendix~\ref{app:study-guides} for the full pre- and post-study interview guides), and an optional preference-video labeling task as part of our technical evaluations (see Section~\ref{sec:eval-representations}).

\subsubsection{\textbf{Data Analysis}}
Given our small sample sizes ($n = 8$ and $n = 7$ per condition), we compared per-participant measures between conditions using two-sided Mann-Whitney $U$ tests. To examine temporal changes across sessions and differences between conditions, we fit linear mixed-effects models with per-participant random effects, using participant-level bootstraps for inference. Qualitative data from pre- and post-study interviews and open-ended post-session responses were analyzed using reflexive thematic analysis~\cite{braun2019reflecting}.

\subsection{Results}
\label{sec:field-study-results}

\subsubsection{\textbf{\system{} promoted more reflective use of recommendation feeds without sacrificing their casual nature}}
\begin{figure}[htbp]
    \centering
    \includegraphics[width=.7\columnwidth]{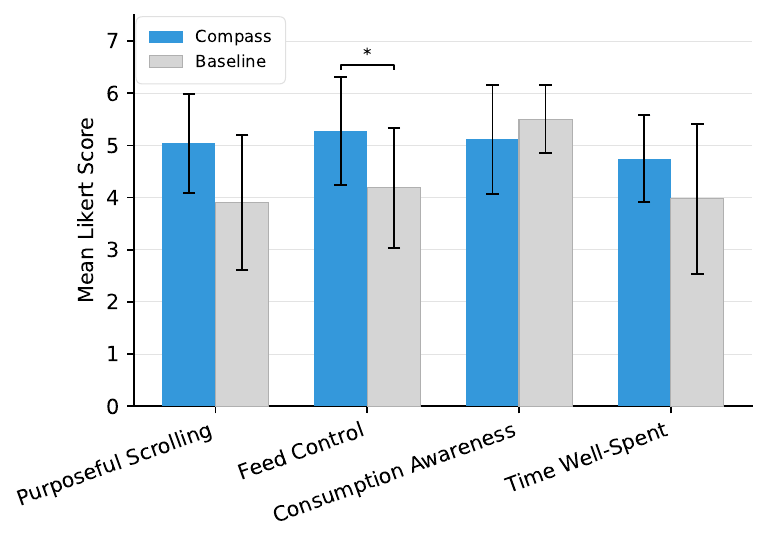}
    \caption{%
    {\normalfont \system{} participants rated Purposeful Scrolling, Feed Control, and Time Well-Spent higher than baseline participants, with no difference in Consumption Awareness.}
    {\normalfont \textbf{Feed Control reached statistical significance ($^*p < .05$).}\strut}
    }
    \label{study-likert}
\end{figure}

Participants using \system{} consistently described feeling more reflective about their social media feeds.
This was supported by during-session self-report data, where \system{} users rated themselves higher than baseline users on scrolling purposefulness and time 
well spent, though neither difference reached significance (Figure~\ref{study-likert}).
In post-study interviews, participants described feeling more mindful of what they were watching, more aware of time spent, and more present during consumption.

Importantly, we find that this added reflection did not come at the cost of the casual nature of the experience. Participants described \system{} as easy to use and non-intrusive, with C3 noting it was \textit{``very user friendly,''} something even their \textit{``little sister or grandma would be able to pick up fairly quickly.''} Comparing \system{} to their normal browsing experience, C2 (see Figure~\ref{usage-timeline-focused} for their interaction timeline) observed that \textit{``for the most part, it was the same, which I think was good---it wasn't taking over too much of the experience.''} 
This was further supported by collected SUS scores, with both \system{} ($84.1 \pm 9.6$) and the baseline ($83.2 \pm 8.6$) falling in the ``Excellent'' range~\cite{bangor2008empirical} and no significant difference between conditions (see Appendix Figure~\ref{sus-per-item} for per-item breakdown), suggesting that adding a reflection layer did not compromise usability.

\begin{figure}[htbp]
    \centering
    \includegraphics[width=\columnwidth]{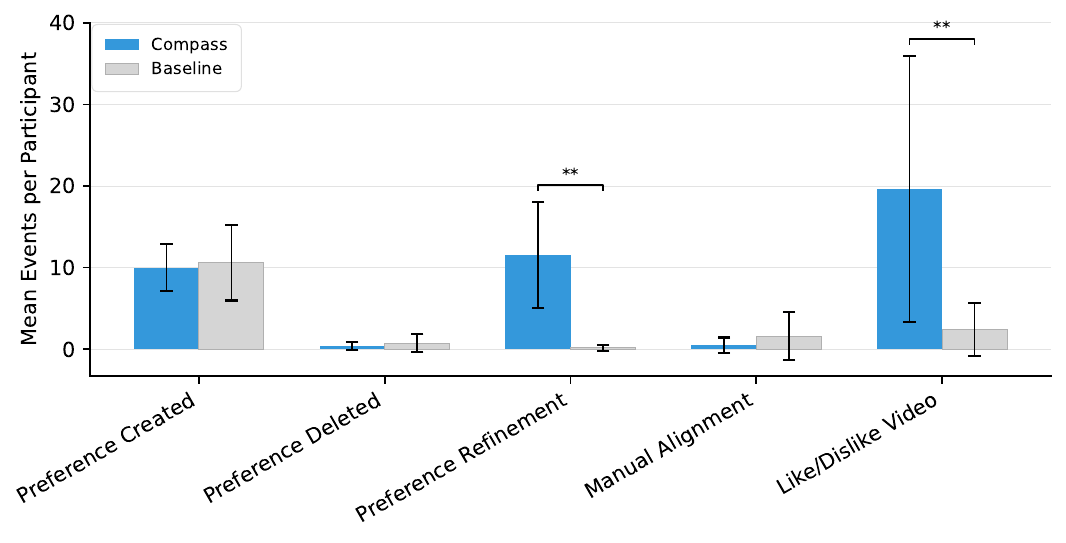}
    \caption{%
    {\normalfont Mean events per participant across both conditions. \system{} participants engaged in significantly more preference refinement and like/dislike interactions, whereas preference creation, preference deletion, and manual alignment were comparable across conditions. ($^{**}p < .01$).}}
    \label{event-counts}
\end{figure}

\textbf{In-situ reflections enabled more contextual and frequent preference refinement, while standalone controls risked losing user awareness}. As shown in Figure~\ref{event-counts}, we observed comparable patterns in preference creation and deletion across conditions, suggesting similar intent to shape the feed. Preference creation was largely reactive and context-driven rather than planned upfront: participants wrote preferences in response to content appearing in their feed (e.g., ``Malcolm in the Middle'', ``Beatles History'') or prompted by events in their daily lives (e.g., ``New York Restaurants'', ``Champions League''). \textit{Avoid} preferences tended to target unwanted content categories such as AI-generated videos, low-quality content, and violent content. 
See Appendix Table~\ref{tab:clusters-k15} for the full set of preference themes identified across participants.

In contrast, a striking difference emerged in preference refinement.
We found that \system{} participants made significantly more refinements than baseline participants (11.50 ± 6.50 vs. 0.14 ± 0.38 per participant, $p = .001$), with baseline participants making only a single refinement in total.
This difference is best explained by where interactions occurred. 
\revise{In \system{}, 73 of 92 refinements 
happened directly in the feed rather than through the preference detail page, and 18 of 80 preferences were also added directly from the feed (see Appendix Tables~\ref{tab:interactive-prompts} and ~\ref{tab:refinement-breakdown}).}
These findings suggest that \system{} enabled easier expression and refinement of preferences without interrupting the browsing flow.
\revise{In interviews, participants further explained this preference: C2 described navigating to the preference detail page as \textit{``a little bit more abrasive\ldots whereas more streamlined when it's coming to you in the feed,''} while C8 noted that having in-feed preference suggestions made it seamless to try them out, and provided a natural starting point for further refinement.}

Without in-feed reflections, refining preferences required more deliberate effort from baseline participants.
\revise{As Figure~\ref{usage-timeline-focused} shows, their session timelines were typically sparse, with activity concentrated in only a few clusters.}
In the most extreme case, B1 forgot the system existed entirely, describing it as \textit{``out of sight, out of mind,''} highlighting that controls requiring users to leave the feed risk being overlooked in the context of casual social media use.
We also observed that most baseline participants struggled to understand how the system interpreted their preferences and whether it was working as intended, due to its lack of transparency features.
As B5 put it, \textit{``I kind of felt like I wasn't grasping it\ldots I sometimes didn't know if I was being too vague in my preferences, or if I needed to be more general with the topics.''}

\subsubsection{\textbf{\system{} achieved better feed alignment through continuous background actions rather than relying on manual user effort}} 
\revise{We found that \system{} users saw significantly more content aligned with their \textit{See More} preferences ($49.3\% \pm 19.0$ vs.\ $20.3\% \pm 14.9$; $p{=}.009$) and devoted a significantly larger share of their watch time to it ($58.7\% \pm 21.4$ vs.\ $21.0\% \pm 11.6$; $p{=}.001$). Nevertheless, we did not observe significant differences in feed alignment and watch time for \textit{Avoid} preferences across two conditions (See Appendix Table~\ref{tab:between-condition-levels} for details).
Additionally, this advantage was front-loaded rather than accumulating over sessions.
We noticed that \system{} feeds began with high alignment and remained stable.
In contrast, baseline feeds improved more rapidly from a much lower starting point, gaining 6.26\% more alignment per session than \system{} feeds
(95\% CI $[2.17, 10.17]$, $p_{\mathrm{boot}} = .014$; Appendix Table~\ref{tab:between-condition-trends}).
Even so, baseline feeds were still less aligned in the end.}


Beyond these aggregate alignment levels, participants also experienced \system{} as adapting to their preferences over time, after an initial learning phase.
C6 noted that early on there was \textit{``a lot of irrelevant content''} where the system would \textit{``misalign a video with a specific preference,''} 
but that by the second or third session their feed got \textit{``dramatically better and became full of videos aligned with my preferences,''} with little content they wanted to avoid.
Additionally, we found that \system{} users became less likely to reject in-situ reflections across sessions ($-3.20$ pts/session, 95\% CI $[-5.75, -0.99]$, $p_{\mathrm{boot}} = .001$; Appendix Table~\ref{tab:notification-outcome-slopes}), indicating that \system{} provided more accurate suggestions over time.

This difference in feed alignment can be explained by the number of actions taken to align the feed: \system{} participants benefited from significantly more background simulations than baseline participants (averaging $21.5 \pm 9.1$ vs.\ $5.9 \pm 2.4$; $p = .002$), driven by \system{}'s proactive feed monitoring, as well as 107 total direct video injections exclusive to \system{} by design.
Interestingly, despite seeing less aligned feeds, baseline participants did not remember to manually align their feeds, averaging fewer than 2 manual alignment triggers (Figure~\ref{event-counts}), consistent with our earlier finding that controls requiring users to leave the feed see substantially reduced engagement.

\textbf{Participants appreciated that \system{} balanced intentional 
feed control with the engaging, discovery-driven experience 
that short-form video platforms provide.} 
\system{} users reported significantly higher perceived feed control than baseline users ($5.28 \pm 1.04$ vs.\ $4.19 \pm 1.15$; $p = .032$; Figure~\ref{study-likert}).
As C3 described, 
\system{} \textit{``still gives you that nice dopamine rush  of finding something good, but it's something you are interested in, not just something random.''} Using the Vendi 
Score~\cite{friedman2022vendi} to measure the content diversity in a feed, we found no significant difference across conditions, suggesting that \system{} improves alignment without reducing content diversity. In some cases, \system{} was able to surface content that participants felt would never have appeared. C3, whose feed was dominated by classic rock and guitar content, was skeptical that electric music 
production would ever surface on its own, yet after setting a \textit{See More} preference, it began appearing, which they described as \textit{``a big help.''} 
However, some participants wished for more granular control of preference alignment.
For instance, C1 expected preferences to surface content in high volume and distributed evenly across all preferences, noting \textit{``I thought I would see all of those pretty equally, and I just feel like it didn't.''} These point to the challenge
of working with a recommendation system that itself offers limited transparency or control.


\section{Technical Evaluation}
\label{sec:eval-representations}
\subsection{Methods}

A fundamental component within \system{} supports matching videos within users' feeds to their reflective preferences. 
Recall that \system{} represents both preferences and videos through a similar three-step process: collecting available inputs (e.g., names, example videos for preferences; metadata, keyframes for videos), using an LLM to translate them into semantically richer descriptions, and finally, embedding these descriptions for efficient comparison.
In this section, we evaluate this approach against baselines to demonstrate that it can reliably support real-time relevance matching. Specifically, we answer the following three questions.

\revise{\begin{itemize}
    \item \textbf{RQ1} Which inputs should we use for video representation, balancing accuracy and latency?
    \item \textbf{RQ2} Does using example videos in the preference representation improve matching accuracy?
    \item \textbf{RQ3} Does the LLM expansion step add enough signal to justify its latency, a key concern in our real-time setting?
\end{itemize}}

\textbf{Representation Ablations.}
Motivated by these research questions, we consider four \emph{video representations}: \textit{Meta}, \textit{Meta${\rightarrow}$LLM}, \textit{Frames${\rightarrow}$LLM}, and \textit{Meta+Frames${\rightarrow}$LLM}.
We exclude a keyframes-only representation without LLM synthesis, since our text-based embedding model cannot encode visual inputs. 
We also consider four \emph{preference representations}: \textit{Name} and \textit{Name${\rightarrow}$LLM}, each with $N = 0, 1, 2$ example videos.
We cap the number of examples at two due to limited positive examples per preference in our dataset.
We include an LLM-as-Judge condition that classifies each preference-video pair from all available information.
It approximates an upper bound on performance but incurs the highest latency, making it impractical to deploy in our setting. We report ROC-AUC and $F1$, with thresholds selected to maximize $F1$.

\begin{table}[t]
\centering
\small
\setlength{\tabcolsep}{5pt}
\begin{tabular}{@{}llcc@{}}
\toprule
\textbf{Preference Repr} & \textbf{Video Repr} & \textbf{ROC-AUC} & \textbf{Best F1} \\
\midrule
Name                        & Meta                          & 0.857 & 0.719 \\
Name                        & Meta${\rightarrow}$LLM     & 0.886  & 0.720 \\
Name                        & Frames${\rightarrow}$LLM   & 0.881  & 0.713 \\
Name                        & Meta+Frames${\rightarrow}$LLM & 0.904  & 0.762\\
Name${\rightarrow}$LLM   & Meta                          & 0.799  & 0.635\\
Name${\rightarrow}$LLM   & Meta${\rightarrow}$LLM     & 0.831 & 0.644 \\
Name${\rightarrow}$LLM   & Frames${\rightarrow}$LLM   & 0.820  & 0.651\\
\rowcolor{blue!8}
Name${\rightarrow}$LLM   & Meta+Frames${\rightarrow}$LLM & 0.842  & 0.675 \\
\midrule
\multicolumn{4}{@{}l@{}}{\textit{Include 1 video example in preference representation}} \\[2pt]

Name+Ex                     & Meta+Frames${\rightarrow}$LLM & 0.904 & 0.731 \\
\rowcolor{blue!8}
Name+Ex${\rightarrow}$LLM & Meta+Frames${\rightarrow}$LLM & 0.878 & 0.675  \\
\midrule
\multicolumn{4}{@{}l@{}}{\textit{Include 2 video examples in preference representation}} \\[2pt]
Name+Ex                     & Meta+Frames${\rightarrow}$LLM & \textbf{0.913}  & \textbf{0.792} \\
\rowcolor{blue!8}
Name+Ex${\rightarrow}$LLM & Meta+Frames${\rightarrow}$LLM & 0.876 & 0.731  \\
\midrule
\multicolumn{4}{@{}l@{}}{\textit{LLM-as-Judge: Prompt LLM once per preference-video pair}} \\[2pt]
\multicolumn{2}{@{}l}{0 video examples} & {---}  & 0.753  \\
\multicolumn{2}{@{}l}{1 video example} & {---}  & 0.748        \\
\multicolumn{2}{@{}l}{2 video examples} & {---}  & 0.744          \\
\bottomrule
\end{tabular}
\vspace{2pt}
\caption{%
    \textbf{We compare different ways of representing preferences and videos before embedding them for similarity matching, evaluated on a test dataset of $N_{\text{eval}} = 408$ preference-video pairs.}
    {\normalfont \emph{Preference Repr}: the preference name alone, or expanded into a topic list via LLM (\textit{Name${\rightarrow}$LLM}), optionally augmented with different numbers of example videos.}
    {\normalfont \emph{Video Repr}: the video's metadata, keyframes, or both, optionally synthesized into a description via LLM (${\rightarrow}$LLM).}
    {\normalfont We also include a LLM-as-judge condition, where the LLM is prompted once per preference-video pair using all available information.}
    {\normalfont We report ROC-AUC and the best F1 across cosine-similarity thresholds for each condition. As LLM-as-Judge condition offers a discrete prediction, ROC-AUC is not applicable. \textbf{Bold}: best value per column across all conditions. \colorbox{blue!8}{Blue shaded rows} indicate the configurations used by \system{}. See Appendix Table~\ref{tab:ablations-full} for complete results.\strut}
}
\label{tab:ablations-compact}
\vspace{-2.5em}
\end{table}

\textbf{Labeled Dataset.}
We construct a ground truth dataset consisting of preference-video pairs labeled \textit{relevant} or \textit{irrelevant}. After the field study concluded, we asked participants to label videos from their own feeds against preferences they had written.
Since most videos are irrelevant to a given preference, we stratified sampling by similarity score to cover a range of relevance levels.
We excluded preferences with insufficient coverage and targeted up to 50 labeled pairs per participant.
Within each preference, videos were presented in random order. Participants saw only the preference name and a video link before making a binary judgment.
The resulting evaluation dataset consists of 408 labeled samples (92 positive, 316 negative) across 36 preferences.
See Appendix~\ref{app:dataset-construction} for details on dataset construction.

\subsection{Results}


\revise{We present our results in Table~\ref{tab:ablations-compact}. See Appendix~\ref{detailed-results} for details. \textbf{(RQ1)} Controlling for the same preference representation, we find that synthesizing both metadata and keyframes with an LLM yields the best video representation, indicating that the two inputs carry complementary signal. \textbf{(RQ2)} Controlling for the same video representation, adding more example videos improves matching, bringing performance close to the LLM-as-Judge upper bound by two examples without its latency. \textbf{(RQ3)} The LLM expansion step improves video representations but offers little benefit for preferences, where embedding the raw name performs at least as well; \system{} nonetheless retains expansion, as the generated subcategories give users a tangible vocabulary to inspect and edit their preferences. Overall, these results show that embedding-based matching is well-suited for real-time feed intervention.}

\section{Discussion}
\subsection{Balancing Casual Browsing with Reflection}

\revise{Our experience designing \system{} surfaced two lessons for systems aimed at supporting reflective use of engagement-driven feeds. First, such systems must preserve the entertaining, low-friction nature of browsing itself; otherwise users lose momentum to keep using the system~\cite{collins2014social, lukoff2018makes}. Our experiments revealed that \system{}'s in-situ reflections enabled participants to more frequently articulate and refine preferences than those in the baseline condition.
Second, observable effects appear to be a precondition for reflective controls to feel worth using. For instance, we noticed that participants in the experiment condition used platform-native controls such as ``likes'' significantly more actively, since clicking these controls triggers \system{} to create a new preference, whereas the same click on its own steers the recommender silently and at a delay~\cite{vaccaro2018illusion, eslami2016first}.}

\revise{However, designing in-feed reflective mechanisms to be lightweight comes with its own limitations. First, because such interactions are intentionally simple and unobtrusive, they cannot easily support more granular control over feed alignment, e.g., surfacing content evenly across reflective preferences, as C1 desired. 
Future work should investigate what kinds of granular controls users might want and how to expose them without breaking the lightweight feel. 
Second, in-feed scaffolding tends to gravitate reflection toward whatever is currently on screen. Even though \system{} includes an overview page meant to surface higher-level patterns, participants predominantly articulated preferences about specific content topics rather than higher-level usage patterns such as content diversity, pacing, or compulsive use. Features such as personalized suggestions of more diverse potential preferences could help users realize the many ways they could express preferences beyond content.}

\subsection{Applying Feed Intervention Techniques beyond Alignment}

\system{} demonstrates a new class of feed intervention techniques for short-form video platforms. While prior work has manipulated text-based feeds~\cite{piccardi2024reranking}, short-form video presents additional challenges: videos are more complex than text posts, and the fast-paced nature of short-form feeds compresses the window for real-time intervention. 
\revise{Given these constraints, \system{} uses lightweight semantic embeddings, which might not yet capture intent- or style-based preferences some participants wanted. Future work should investigate supporting these richer visual features without compromising real-time responsiveness.}
\revise{Additionally, while implementing \system{} as a browser extension constrained our study to desktop, its interaction paradigm and user interface are mobile-friendly. We envision similar approaches generalizing to mobile platforms, where equivalent DOM-level manipulations could be achieved through accessibility APIs~\cite{lu2024interactout, li2017sugilite}.}

To align recommendation feeds with users' reflective preferences, we draw on two technical approaches, both of which have applications beyond feed alignment.
The first approach assumes the algorithm remains responsive to engagement signals and steers it through simulated user behavior. More broadly, the ability to programmatically generate engagement signals beyond users' actual behavior opens up uses across adjacent settings. A user migrating across platforms could use simulated signals to seed a new feed with preferences carried over from their previous platform, rather than starting from scratch~\cite{jamieson2023escaping, fiesler2020moving}. A trusted party could shape someone else's feed on their behalf, such as a parent helping to shape a child's feed, or an adult doing the same for an elderly parent~\cite{chowdhury2023co}. Researchers could use synthetic engagement signals as an experimental probe, systematically varying the input to study how recommendation algorithms respond over time~\cite{sandvig2014auditing, haroon2023auditing}.

The second approach assumes the platform may not act in users' interests, and instead manipulates the feed on the client side without relying on platform cooperation. The same mechanism opens up other ways for users to reclaim control over their feed experience. Outgoing interaction signals could be intercepted before reaching the platform to keep behavioral data local~\cite{erlingsson2014rappor}. Additionally, ads and platform-pushed content that users have no interest in could be filtered out directly, extending the spirit of existing ad blockers to the broader recommendation layer~\cite{mathur2018characterizing, jhaver2023personalizing}.

\section{Limitations}
Our study has several limitations. First, \system{} was developed and evaluated on desktop YouTube Shorts. Although we designed the interface to mirror mobile interaction patterns, it remains an open challenge to extend the prototype to mobile platforms, where short-form video consumption is more prevalent and technical constraints for intervention differ. \revise{Second, our field study had a small per-condition sample ($n=8$ \system{}, $n=7$ baseline), which limits statistical power. Accordingly, we treat our results as suggestive rather than definitive.} Third, our participants were recruited from a university population, limiting the generalizability of our findings to older adults or to those less familiar with short-form video platforms. Future studies with larger and more diverse samples would provide a comprehensive understanding of how users across different backgrounds engage with reflective feed alignment tools. \revise{Finally, while our study offers preliminary evidence of a sustained effect, future longitudinal diary studies can examine continued use beyond the novelty effect and how reflective preferences change over time.}

\section{Conclusion}
We presented \system{}, a system that treats feed alignment, and reflection as a continuous process integrated into everyday social media use, rather than a one-time configuration task. Our work suggests that in-situ interactions during feed consumption can meaningfully close the gap between how engagement-based algorithms shape feeds and what users actually want after reflection. 

\begin{acks}
This research was supported by NSF Award \#2348926. We would like to thank members of the Social Futures Lab at the University of Washington for their helpful feedback throughout the work. We would also like to thank all the study participants who provided us with valuable insights through their use of the system. Lastly, we would like to thank our anonymous reviewers for their insightful feedback on our manuscript. 
\end{acks}

\bibliographystyle{ACM-Reference-Format}
\bibliography{references}

@article{piccardi2024reranking,
  title={Reranking social media feeds: A practical guide for field experiments},
  author={Piccardi, Tiziano and Saveski, Martin and Jia, Chenyan and Hancock, Jeffrey and Tsai, Jeanne L and Bernstein, Michael S},
  journal={arXiv preprint arXiv:2406.19571},
  year={2024}
}

@article{friedman2022vendi,
  title={The vendi score: A diversity evaluation metric for machine learning},
  author={Friedman, Dan and Dieng, Adji Bousso},
  journal={arXiv preprint arXiv:2210.02410},
  year={2022}
}

@techreport{ricks2022youtube,           
  author      = {Ricks, Becca and McCrosky, Jesse},  
  title       = {Does This Button Work? Investigating YouTube's Ineffective User Controls},  
  institution = {Mozilla Foundation},     
  year        = {2022},                   
  month       = sep,                      
  url         = {https://assets.mofoprod.net/network/documents/Mozilla-Report-YouTube-User-Controls.pdf}  
}

@article{jia2024embedding,
  title={Embedding democratic values into social media AIs via societal objective functions},
  author={Jia, Chenyan and Lam, Michelle S and Mai, Minh Chau and Hancock, Jeffrey T and Bernstein, Michael S},
  journal={Proceedings of the ACM on Human-Computer Interaction},
  volume={8},
  number={CSCW1},
  pages={1--36},
  year={2024},
  publisher={ACM New York, NY, USA}
}

@article{kolluri2025alexandria,
  title={Alexandria: A Library of Pluralistic Values for Realtime Re-Ranking of Social Media Feeds},
  author={Kolluri, Akaash and Su, Renn and Jahanbakhsh, Farnaz and Zhao, Dora and Piccardi, Tiziano and Bernstein, Michael S},
  journal={arXiv preprint arXiv:2505.10839},
  year={2025}
}

@article{malki2025bonsai,
  title={Bonsai: Intentional and personalized social media feeds},
  author={Malki, Omar El and Qu{\'e}r{\'e}, Marianne Aubin Le and Monroy-Hern{\'a}ndez, Andr{\'e}s and Ribeiro, Manoel Horta},
  journal={arXiv preprint arXiv:2509.10776},
  year={2025}
}

@article{liu2025understanding,
  title={Understanding decentralized social feed curation on mastodon},
  author={Liu, Yuhan and Song, Emmy and Zhang, Owen Xingjian and Merriman, Jewel and Zhang, Lei and Monroy-Hern{\'a}ndez, Andr{\'e}s},
  journal={Proceedings of the ACM on Human-Computer Interaction},
  volume={9},
  number={7},
  pages={1--25},
  year={2025},
  publisher={ACM New York, NY, USA}
}

@inproceedings{bhargava2019gobo,
  title={Gobo: A system for exploring user control of invisible algorithms in social media},
  author={Bhargava, Rahul and Chung, Anna and Gaikwad, Neil S and Hope, Alexis and Jen, Dennis and Rubinovitz, Jasmin and Sald{\'\i}as-Fuentes, Bel{\'e}n and Zuckerman, Ethan},
  booktitle={Companion publication of the 2019 conference on computer supported cooperative work and social computing},
  pages={151--155},
  year={2019}
}

@article{choi2025designing,
  title={Designing Usable Controls for Customizable Social Media Feeds},
  author={Choi, Frederick and Chandrasekharan, Eshwar},
  journal={arXiv preprint arXiv:2509.19615},
  year={2025}
}

@article{quelle2025bluesky,
  title={Bluesky: Network topology, polarization, and algorithmic curation},
  author={Quelle, Dorian and Bovet, Alexandre},
  journal={PloS one},
  volume={20},
  number={2},
  pages={e0318034},
  year={2025},
  publisher={Public Library of Science San Francisco, CA USA}
}

@article{braun2019reflecting,
  title={Reflecting on reflexive thematic analysis},
  author={Braun, Virginia and Clarke, Victoria},
  journal={Qualitative research in sport, exercise and health},
  volume={11},
  number={4},
  pages={589--597},
  year={2019},
  publisher={Taylor \& Francis}
}

@inproceedings{lukoff2021design,
  title={How the design of youtube influences user sense of agency},
  author={Lukoff, Kai and Lyngs, Ulrik and Zade, Himanshu and Liao, J Vera and Choi, James and Fan, Kaiyue and Munson, Sean A and Hiniker, Alexis},
  booktitle={Proceedings of the 2021 CHI Conference on Human Factors in Computing Systems},
  pages={1--17},
  year={2021}
}

@inproceedings{eslami2015always,
  title={" I always assumed that I wasn't really that close to [her]" Reasoning about invisible algorithms in news feeds},
  author={Eslami, Motahhare and Rickman, Aimee and Vaccaro, Kristen and Aleyasen, Amirhossein and Vuong, Andy and Karahalios, Karrie and Hamilton, Kevin and Sandvig, Christian},
  booktitle={Proceedings of the 33rd annual ACM conference on human factors in computing systems},
  pages={153--162},
  year={2015}
}

@inproceedings{eslami2016first,
  title={First I" like" it, then I hide it: Folk Theories of Social Feeds},
  author={Eslami, Motahhare and Karahalios, Karrie and Sandvig, Christian and Vaccaro, Kristen and Rickman, Aimee and Hamilton, Kevin and Kirlik, Alex},
  booktitle={Proceedings of the 2016 cHI conference on human factors in computing systems},
  pages={2371--2382},
  year={2016}
}

@inproceedings{devito2017algorithms,
  title={" Algorithms ruin everything" \# RIPTwitter, Folk Theories, and Resistance to Algorithmic Change in Social Media},
  author={DeVito, Michael Ann and Gergle, Darren and Birnholtz, Jeremy},
  booktitle={Proceedings of the 2017 CHI conference on human factors in computing systems},
  pages={3163--3174},
  year={2017}
}

@article{carroll2025ctrl,
  title={CTRL-Rec: Controlling recommender systems with natural language},
  author={Carroll, Micah and Foote, Adeline and Feng, Kevin and Williams, Marcus and Dragan, Anca and Knox, W Bradley and Milli, Smitha},
  journal={arXiv preprint arXiv:2510.12742},
  year={2025}
}

@inproceedings{feng2024mapping,
  title={Mapping the Design Space of Teachable Social Media Feed Experiences},
  author={Feng, KJ Kevin and Koo, Xander and Tan, Lawrence and Bruckman, Amy and McDonald, David W and Zhang, Amy X},
  booktitle={Proceedings of the 2024 CHI Conference on Human Factors in Computing Systems},
  pages={1--20},
  year={2024}
}

@techreport{agan2023automating,
  title={Automating automaticity: How the context of human choice affects the extent of algorithmic bias},
  author={Agan, Amanda Y and Davenport, Diag and Ludwig, Jens and Mullainathan, Sendhil},
  year={2023},
  institution={National Bureau of Economic Research}
}

@article{morewedge2023human,
  title={Human bias in algorithm design},
  author={Morewedge, Carey K and Mullainathan, Sendhil and Naushan, Haaya F and Sunstein, Cass R and Kleinberg, Jon and Raghavan, Manish and Ludwig, Jens O},
  journal={Nature Human Behaviour},
  volume={7},
  number={11},
  pages={1822--1824},
  year={2023},
  publisher={Nature Publishing Group UK London}
}

@article{brady2017emotion,
  title={Emotion shapes the diffusion of moralized content in social networks},
  author={Brady, William J and Wills, Julian A and Jost, John T and Tucker, Joshua A and Van Bavel, Jay J},
  journal={Proceedings of the National Academy of Sciences},
  volume={114},
  number={28},
  pages={7313--7318},
  year={2017},
  publisher={National Academy of Sciences}
}

@article{haroon2023auditing,
  title={Auditing YouTube’s recommendation system for ideologically congenial, extreme, and problematic recommendations},
  author={Haroon, Muhammad and Wojcieszak, Magdalena and Chhabra, Anshuman and Liu, Xin and Mohapatra, Prasant and Shafiq, Zubair},
  journal={Proceedings of the national academy of sciences},
  volume={120},
  number={50},
  pages={e2213020120},
  year={2023},
  publisher={National Academy of Sciences}
}

@article{brady2023algorithm,
  title={Algorithm-mediated social learning in online social networks},
  author={Brady, William J and Jackson, Joshua Conrad and Lindstr{\"o}m, Bj{\"o}rn and Crockett, MJ},
  journal={Trends in cognitive sciences},
  volume={27},
  number={10},
  pages={947--960},
  year={2023},
  publisher={Elsevier}
}

@inproceedings{ekstrand2016behaviorism,
  title={Behaviorism is not enough: better recommendations through listening to users},
  author={Ekstrand, Michael D and Willemsen, Martijn C},
  booktitle={Proceedings of the 10th ACM conference on recommender systems},
  pages={221--224},
  year={2016}
}

@inproceedings{yang2025studying,
  title={Studying behavioral addiction by combining surveys and digital traces: A case study of TikTok},
  author={Yang, Cai and Mousavi, Sepehr and Dash, Abhisek and Gummadi, Krishna P and Weber, Ingmar},
  booktitle={Proceedings of the International AAAI Conference on Web and Social Media},
  volume={19},
  pages={2106--2123},
  year={2025}
}

@article{cunningham2025ranking,
  title={Ranking by engagement and non-engagement signals: Learnings from industry},
  author={Cunningham, Tom and Pandey, Sana and Sigerson, Leif and Stray, Jonathan and Allen, Jeff and Barrilleaux, Bonnie and Iyer, Ravi and Kothari, Mohit and Rezaei, Behnam and Kairam, Sanjay and others},
  journal={Annals of the New York Academy of Sciences},
  volume={1551},
  number={1},
  pages={19--32},
  year={2025},
  publisher={Wiley Online Library}
}

@article{cunningham2024we,
  title={What we know about using non-engagement signals in content ranking},
  author={Cunningham, Tom and Pandey, Sana and Sigerson, Leif and Stray, Jonathan and Allen, Jeff and Barrilleaux, Bonnie and Iyer, Ravi and Milli, Smitha and Kothari, Mohit and Rezaei, Behnam},
  journal={arXiv preprint arXiv:2402.06831},
  year={2024}
}

@article{kim2019goalkeeper,
  title={Goalkeeper: Exploring interaction lockout mechanisms for regulating smartphone use},
  author={Kim, Jaejeung and Jung, Hayoung and Ko, Minsam and Lee, Uichin},
  journal={Proceedings of the ACM on Interactive, Mobile, Wearable and Ubiquitous Technologies},
  volume={3},
  number={1},
  pages={1--29},
  year={2019},
  publisher={ACM New York, NY, USA}
}

@inproceedings{kim2016timeaware,
  title={TimeAware: Leveraging framing effects to enhance personal productivity},
  author={Kim, Young-Ho and Jeon, Jae Ho and Choe, Eun Kyoung and Lee, Bongshin and Kim, KwonHyun and Seo, Jinwook},
  booktitle={Proceedings of the 2016 CHI Conference on Human Factors in Computing Systems},
  pages={272--283},
  year={2016}
}

@inproceedings{collins2014social,
  title={Social networking use and RescueTime: the issue of engagement},
  author={Collins, Emily IM and Cox, Anna L and Bird, Jon and Harrison, Daniel},
  booktitle={Proceedings of the 2014 ACM International Joint Conference on Pervasive and Ubiquitous Computing: Adjunct Publication},
  pages={687--690},
  year={2014}
}

@inproceedings{choe2015sleeptight,
  title={SleepTight: low-burden, self-monitoring technology for capturing and reflecting on sleep behaviors},
  author={Choe, Eun Kyoung and Lee, Bongshin and Kay, Matthew and Pratt, Wanda and Kientz, Julie A},
  booktitle={Proceedings of the 2015 ACM international joint conference on pervasive and ubiquitous computing},
  pages={121--132},
  year={2015}
}

@article{nelson1981theoretical,
  title={Theoretical explanations for reactivity in self-monitoring},
  author={Nelson, Rosemery O and Hayes, Steven C},
  journal={Behavior Modification},
  volume={5},
  number={1},
  pages={3--14},
  year={1981},
  publisher={Sage Publications Sage CA: Thousand Oaks, CA}
}

@inproceedings{consolvo2008flowers,
  title={Flowers or a robot army? Encouraging awareness \& activity with personal, mobile displays},
  author={Consolvo, Sunny and Klasnja, Predrag and McDonald, David W and Avrahami, Daniel and Froehlich, Jon and LeGrand, Louis and Libby, Ryan and Mosher, Keith and Landay, James A},
  booktitle={Proceedings of the 10th international conference on Ubiquitous computing},
  pages={54--63},
  year={2008}
}

@article{lukoff2018makes,
  title={What makes smartphone use meaningful or meaningless?},
  author={Lukoff, Kai and Yu, Cissy and Kientz, Julie and Hiniker, Alexis},
  journal={Proceedings of the ACM on Interactive, Mobile, Wearable and Ubiquitous Technologies},
  volume={2},
  number={1},
  pages={1--26},
  year={2018},
  publisher={ACM New York, NY, USA}
}

@article{kleinberg2024challenge,
  title={The challenge of understanding what users want: Inconsistent preferences and engagement optimization},
  author={Kleinberg, Jon and Mullainathan, Sendhil and Raghavan, Manish},
  journal={Management science},
  volume={70},
  number={9},
  pages={6336--6355},
  year={2024},
  publisher={INFORMS}
}

@article{jhaver2023personalizing,
  title={Personalizing content moderation on social media: User perspectives on moderation choices, interface design, and labor},
  author={Jhaver, Shagun and Zhang, Alice Qian and Chen, Quan Ze and Natarajan, Nikhila and Wang, Ruotong and Zhang, Amy X},
  journal={Proceedings of the ACM on Human-Computer Interaction},
  volume={7},
  number={CSCW2},
  pages={1--33},
  year={2023},
  publisher={ACM New York, NY, USA}
}

@inproceedings{wang2025end,
  title={End user authoring of personalized content classifiers: comparing example labeling, rule writing, and LLM prompting},
  author={Wang, Leijie and Yurechko, Kathryn and Dani, Pranati and Chen, Quan Ze and Zhang, Amy X},
  booktitle={Proceedings of the 2025 CHI Conference on Human Factors in Computing Systems},
  pages={1--21},
  year={2025}
}

@inproceedings{vaccaro2018illusion,
  title={The illusion of control: Placebo effects of control settings},
  author={Vaccaro, Kristen and Huang, Dylan and Eslami, Motahhare and Sandvig, Christian and Hamilton, Kevin and Karahalios, Karrie},
  booktitle={Proceedings of the 2018 CHI Conference on Human Factors in Computing Systems},
  pages={1--13},
  year={2018}
}

@article{shen2021everyday,
  title={Everyday algorithm auditing: Understanding the power of everyday users in surfacing harmful algorithmic behaviors},
  author={Shen, Hong and DeVos, Alicia and Eslami, Motahhare and Holstein, Kenneth},
  journal={Proceedings of the ACM on Human-Computer Interaction},
  volume={5},
  number={CSCW2},
  pages={1--29},
  year={2021},
  publisher={ACM New York, NY, USA}
}

@article{tang2025interactive,
  title={Interactive Recommendation Agent with Active User Commands},
  author={Tang, Jiakai and Luo, Yujie and Xi, Xunke and Sun, Fei and Feng, Xueyang and Dai, Sunhao and Yi, Chao and Chen, Dian and Gao, Zhujin and Li, Yang and others},
  journal={arXiv preprint arXiv:2509.21317},
  year={2025}
}

@inproceedings{lewis1995sequential,
  title={A sequential algorithm for training text classifiers: Corrigendum and additional data},
  author={Lewis, David D},
  booktitle={Acm Sigir Forum},
  volume={29},
  number={2},
  pages={13--19},
  year={1995},
  organization={ACM New York, NY, USA}
}

@inproceedings{seung1992query,
  title={Query by committee},
  author={Seung, H Sebastian and Opper, Manfred and Sompolinsky, Haim},
  booktitle={Proceedings of the fifth annual workshop on Computational learning theory},
  pages={287--294},
  year={1992}
}

@inproceedings{li2025beyond,
  title={Beyond Explicit and Implicit: How Users Provide Feedback to Shape Personalized Recommendation Content},
  author={Li, Wenqi and Kuo, Jui-Ching and Sheng, Manyu and Zhang, Pengyi and Wu, Qunfang},
  booktitle={Proceedings of the 2025 CHI Conference on Human Factors in Computing Systems},
  pages={1--17},
  year={2025}
}

@article{kumar2023watch,
  title={Watch your language: large language models and content moderation},
  author={Kumar, Deepak and AbuHashem, Yousef and Durumeric, Zakir},
  journal={arXiv preprint arXiv:2309.14517},
  year={2023}
}

@misc{youtube_press_2026,
  author       = {{YouTube}},
  title        = {YouTube for Press},
  year         = {2026},
  howpublished = {\url{https://blog.youtube/press/}},
  note         = {Accessed: 2026-03-21}
}

@inproceedings{covington2016deep,
  title={Deep neural networks for youtube recommendations},
  author={Covington, Paul and Adams, Jay and Sargin, Emre},
  booktitle={Proceedings of the 10th ACM conference on recommender systems},
  pages={191--198},
  year={2016}
}

@inproceedings{zannettou2024analyzing,
  title={Analyzing user engagement with TikTok's short format video recommendations using data donations},
  author={Zannettou, Savvas and Nemes-Nemeth, Olivia and Ayalon, Oshrat and Goetzen, Angelica and Gummadi, Krishna P and Redmiles, Elissa M and Roesner, Franziska},
  booktitle={Proceedings of the 2024 CHI Conference on Human Factors in Computing Systems},
  pages={1--16},
  year={2024}
}

@book{kahneman2011thinking,
  title={Thinking, fast and slow},
  author={Kahneman, Daniel},
  year={2011},
  publisher={macmillan}
}

@inproceedings{vera2025they,
  title={" They've Over-Emphasized That One Search": Controlling Unwanted Content on TikTok's For You Page},
  author={Vera, Julie A and Ghosh, Sourojit},
  booktitle={Proceedings of the 2025 CHI Conference on Human Factors in Computing Systems},
  pages={1--8},
  year={2025}
}

@inproceedings{vombatkere2024tiktok,
  title={Tiktok and the art of personalization: investigating exploration and exploitation on social media feeds},
  author={Vombatkere, Karan and Mousavi, Sepehr and Zannettou, Savvas and Roesner, Franziska and Gummadi, Krishna P},
  booktitle={Proceedings of the ACM Web Conference 2024},
  pages={3789--3797},
  year={2024}
}

@inproceedings{eslami2019user,
  title={User attitudes towards algorithmic opacity and transparency in online reviewing platforms},
  author={Eslami, Motahhare and Vaccaro, Kristen and Lee, Min Kyung and Elazari Bar On, Amit and Gilbert, Eric and Karahalios, Karrie},
  booktitle={Proceedings of the 2019 CHI Conference on Human Factors in Computing Systems},
  pages={1--14},
  year={2019}
}

@inproceedings{baughan2022don,
  title={“I don’t even remember what I read”: How design influences dissociation on social media},
  author={Baughan, Amanda and Zhang, Mingrui Ray and Rao, Raveena and Lukoff, Kai and Schaadhardt, Anastasia and Butler, Lisa D and Hiniker, Alexis},
  booktitle={Proceedings of the 2022 CHI conference on human factors in computing systems},
  pages={1--13},
  year={2022}
}

@article{popowski2026social,
  title={Social Media Feed Elicitation},
  author={Popowski, Lindsay and Wu, Xiyuan and Zhu, Charlotte and Piccardi, Tiziano and Bernstein, Michael S},
  journal={arXiv preprint arXiv:2602.18594},
  year={2026}
}

@article{guess2023social,
  title={How do social media feed algorithms affect attitudes and behavior in an election campaign?},
  author={Guess, Andrew M and Malhotra, Neil and Pan, Jennifer and Barber{\'a}, Pablo and Allcott, Hunt and Brown, Taylor and Crespo-Tenorio, Adriana and Dimmery, Drew and Freelon, Deen and Gentzkow, Matthew and others},
  journal={Science},
  volume={381},
  number={6656},
  pages={398--404},
  year={2023},
  publisher={American Association for the Advancement of Science}
}

@misc{gupta2021feedback,
  title={Incorporating More Feedback Into News Feed Ranking},
  author={Gupta, Aastha},
  year={2021},
  month={apr},
  howpublished={Meta Newsroom},
  note={Accessed: 2026-03-26},
  url={https://about.fb.com/news/2021/04/incorporating-more-feedback-into-news-feed-ranking/}
}

@misc{meta2022customize,
  title={New Ways to Customize Your Facebook Feed},
  author={{Meta}},
  year={2022},
  month={oct},
  howpublished={Meta Newsroom},
  note={Accessed: 2026-03-26},
  url={https://about.fb.com/news/2022/10/new-ways-to-customize-your-facebook-feed/}
}

@misc{goodrow2021recommendation,
  title={On YouTube’s recommendation system},
  author={Goodrow, Cristos},
  year={2021},
  month={sep},
  howpublished={YouTube Blog},
  note={Accessed: 2026-03-26},
  url={https://blog.youtube/inside-youtube/on-youtubes-recommendation-system/}
}

@inproceedings{hsu2020awareness,
  title={Awareness, navigation, and use of feed control settings online},
  author={Hsu, Silas and Vaccaro, Kristen and Yue, Yin and Rickman, Aimee and Karahalios, Karrie},
  booktitle={Proceedings of the 2020 CHI Conference on Human Factors in Computing Systems},
  pages={1--13},
  year={2020}
}

@inproceedings{liu2011analyzing,
  title={Analyzing facebook privacy settings: user expectations vs. reality},
  author={Liu, Yabing and Gummadi, Krishna P and Krishnamurthy, Balachander and Mislove, Alan},
  booktitle={Proceedings of the 2011 ACM SIGCOMM conference on Internet measurement conference},
  pages={61--70},
  year={2011}
}

@inproceedings{rader2018explanations,
  title={Explanations as mechanisms for supporting algorithmic transparency},
  author={Rader, Emilee and Cotter, Kelley and Cho, Janghee},
  booktitle={Proceedings of the 2018 CHI conference on human factors in computing systems},
  pages={1--13},
  year={2018}
}

@inproceedings{boonprakong2025hci,
  title={How do HCI researchers study cognitive biases? A scoping review},
  author={Boonprakong, Nattapat and Tag, Benjamin and Goncalves, Jorge and Dingler, Tilman},
  booktitle={Proceedings of the 2025 CHI Conference on Human Factors in Computing Systems},
  pages={1--20},
  year={2025}
}

@inproceedings{rader2015understanding,
  title={Understanding user beliefs about algorithmic curation in the Facebook news feed},
  author={Rader, Emilee and Gray, Rebecca},
  booktitle={Proceedings of the 33rd annual ACM conference on human factors in computing systems},
  pages={173--182},
  year={2015}
}

@inproceedings{carroll2022estimating,
  title={Estimating and penalizing induced preference shifts in recommender systems},
  author={Carroll, Micah D and Dragan, Anca and Russell, Stuart and Hadfield-Menell, Dylan},
  booktitle={International Conference on Machine Learning},
  pages={2686--2708},
  year={2022},
  organization={PMLR}
}

@incollection{samuelson2024note,
  title={A note on the pure theory of consumer's behaviour},
  author={Samuelson, Paul A},
  booktitle={The Foundations of Price Theory Vol 4},
  pages={101--116},
  year={2024},
  publisher={Routledge}
}

@inproceedings{lu2024interactout,
  title={InteractOut: leveraging interaction proxies as input manipulation strategies for reducing smartphone overuse},
  author={Lu, Tao and Zheng, Hongxiao and Zhang, Tianying and Xu, Xuhai “Orson” and Guo, Anhong},
  booktitle={Proceedings of the 2024 CHI conference on human factors in computing systems},
  pages={1--19},
  year={2024}
}

@inproceedings{li2017sugilite,
  title={SUGILITE: creating multimodal smartphone automation by demonstration},
  author={Li, Toby Jia-Jun and Azaria, Amos and Myers, Brad A},
  booktitle={Proceedings of the 2017 CHI conference on human factors in computing systems},
  pages={6038--6049},
  year={2017}
}

@article{jamieson2023escaping,
  title={Escaping the Walled Garden? User Perspectives of Control in Data Portability for Social Media},
  author={Jamieson, Jack and Yamashita, Naomi},
  journal={Proceedings of the ACM on Human-Computer Interaction},
  volume={7},
  number={CSCW2},
  pages={1--27},
  year={2023},
  publisher={ACM New York, NY, USA}
}

@article{fiesler2020moving,
  title={Moving across lands: Online platform migration in fandom communities},
  author={Fiesler, Casey and Dym, Brianna},
  journal={Proceedings of the ACM on Human-Computer Interaction},
  volume={4},
  number={CSCW1},
  pages={1--25},
  year={2020},
  publisher={ACM New York, NY, USA}
}

@inproceedings{chowdhury2023co,
  title={Co-Designing with Early Adolescents: Understanding Perceptions of and Design Considerations for Tech-Based Mediation Strategies that Promote Technology Disengagement},
  author={Chowdhury, Ananta and Bunt, Andrea},
  booktitle={Proceedings of the 2023 CHI Conference on Human Factors in Computing Systems},
  pages={1--16},
  year={2023}
}

@article{sandvig2014auditing,
  title={Auditing algorithms: Research methods for detecting discrimination on internet platforms},
  author={Sandvig, Christian and Hamilton, Kevin and Karahalios, Karrie and Langbort, Cedric},
  journal={Data and discrimination: converting critical concerns into productive inquiry},
  volume={22},
  number={2014},
  pages={4349--4357},
  year={2014}
}

@inproceedings{erlingsson2014rappor,
  title={Rappor: Randomized aggregatable privacy-preserving ordinal response},
  author={Erlingsson, {\'U}lfar and Pihur, Vasyl and Korolova, Aleksandra},
  booktitle={Proceedings of the 2014 ACM SIGSAC conference on computer and communications security},
  pages={1054--1067},
  year={2014}
}

@inproceedings{mathur2018characterizing,
  title={Characterizing the use of $\{$Browser-Based$\}$ blocking extensions to prevent online tracking},
  author={Mathur, Arunesh and Vitak, Jessica and Narayanan, Arvind and Chetty, Marshini},
  booktitle={Fourteenth symposium on usable privacy and security (SOUPS 2018)},
  pages={103--116},
  year={2018}
}

@article{brooke1996sus,
  title={SUS-A quick and dirty usability scale},
  author={Brooke, John and others},
  journal={Usability evaluation in industry},
  volume={189},
  number={194},
  pages={4--7},
  year={1996},
  publisher={London, England}
}

@article{bangor2008empirical,
  title={An empirical evaluation of the system usability scale},
  author={Bangor, Aaron and Kortum, Philip T and Miller, James T},
  journal={Intl. Journal of Human--Computer Interaction},
  volume={24},
  number={6},
  pages={574--594},
  year={2008},
  publisher={Taylor \& Francis}
}

@misc{stray2021beyond,
  author       = {Stray, Jonathan},
  title        = {Beyond Engagement: Aligning Algorithmic Recommendations with Prosocial Goals},
  year         = {2021},
  howpublished = {Partnership on AI},
  url          = {https://partnershiponai.org/beyond-engagement-aligning-algorithmic-recommendations-with-prosocial-goals/},
  note         = {Accessed: 2026}
}

@techreport{pew2025socialmedia,
  author      = {{Pew Research Center}},
  title       = {Americans' Social Media Use 2025},
  institution = {Pew Research Center},
  year        = {2025},
  month       = nov,
  url         = {https://www.pewresearch.org/internet/2025/11/20/americans-social-media-use-2025/}
}

@incollection{ruiz2024design,
  title={Design frictions on social media: Balancing reduced mindless scrolling and user satisfaction},
  author={Ruiz, Nicolas and Molina Le{\'o}n, Gabriela and Heuer, Hendrik},
  booktitle={Proceedings of Mensch und Computer 2024},
  pages={442--447},
  year={2024}
}

@inproceedings{mildner2021ethical,
  title={Ethical user interfaces: Exploring the effects of dark patterns on facebook},
  author={Mildner, Thomas and Savino, Gian-Luca},
  booktitle={Extended Abstracts of the 2021 CHI Conference on Human Factors in Computing Systems},
  pages={1--7},
  year={2021}
}

@inproceedings{tan2025curious,
  title={Curious shorts: Curiosity-driven exploration and learning on short-form video platforms},
  author={Tan, Felicia Fang-Yi and Ram, Ashwin and Messerschmidt, Moritz Alexander and Dissanayake, Hasini Amanda and Nanayakkara, Suranga},
  booktitle={Proceedings of the 2025 CHI Conference on Human Factors in Computing Systems},
  pages={1--22},
  year={2025}
}

\appendix

\section{Field Study}

\subsection{Participants}
\label{appendix:participants}

Table~\ref{tab:participants} provides complete demographic and social media usage information for all 15 participants in our field study. These same participants also contributed to the dataset construction process for our technical evaluation (see Appendix~\ref{app:dataset-construction}).

\subsection{Usability}

\begin{figure}[h]
    \centering
    \includegraphics[width=\columnwidth]{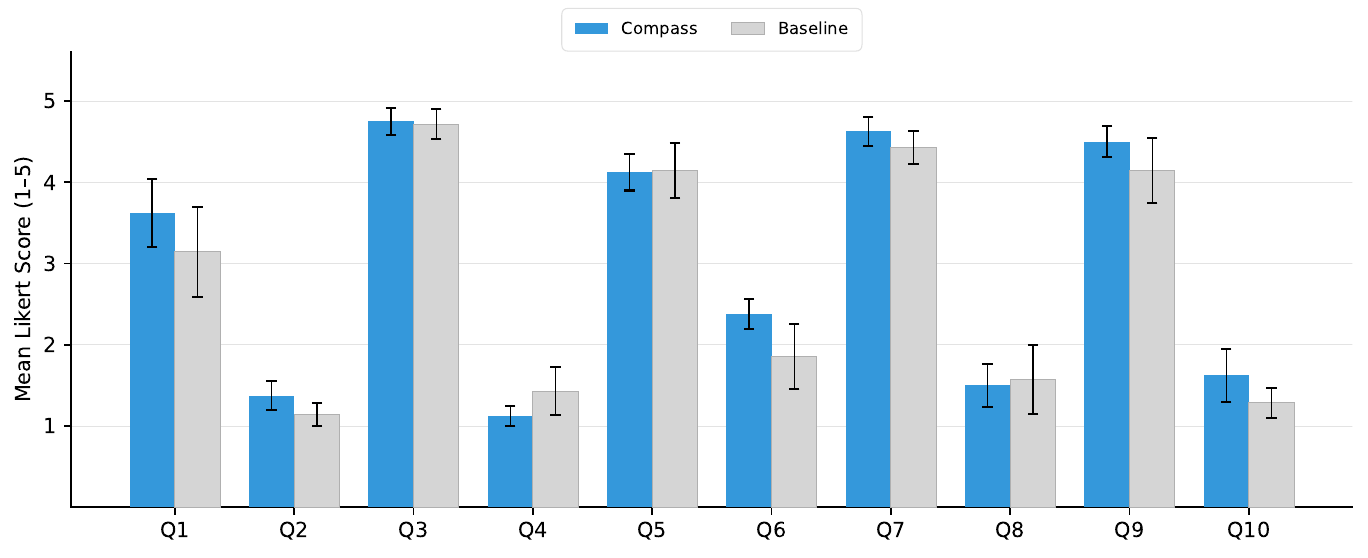}
    \caption{%
      \textbf{Per-item mean SUS responses comparing \system{} and Baseline on the 1--5 Likert scale across all 10 questionnaire items.}
      {\normalfont \emph{Items}: Q1 (would use frequently), Q2 (unnecessarily complex), Q3 (easy to use), Q4 (need technical support), Q5 (well integrated), Q6 (too much inconsistency), Q7 (learn quickly), Q8
  (cumbersome to use), Q9 (felt confident), Q10 (needed to learn a lot).}
      {\normalfont \emph{Polarity}: odd-numbered items are positively worded (higher scores indicate better usability); even-numbered items are negatively worded (lower scores indicate better usability).}
      {\normalfont No per-item difference reached statistical significance.\strut}
  }
    \label{sus-per-item}
\end{figure}

The aggregate SUS scores reported in Section~\ref{sec:field-study-results} (\system{}: 84.1, Baseline: 83.2) place both systems in the ``Excellent'' usability range~\cite{bangor2008empirical}. Figure~\ref{sus-per-item} breaks these scores down by item. Both conditions rated the systems highly on ease of use (Q3: 4.75 vs.\ 4.71), learnability (Q7: 4.62 vs.\ 4.43), and low need for technical support (Q4: 1.12 vs.\ 1.43), and similarly low on cumbersomeness (Q8: 1.50 vs.\ 1.57) and prior learning required (Q10: 1.62 vs.\ 1.29). \system{} participants rated the system slightly higher on willingness to use frequently (Q1: 3.62 vs.\ 3.14) and confidence using it (Q9: 4.50 vs.\ 4.14), while reporting marginally more perceived inconsistency (Q6: 2.38 vs.\ 1.86), plausibly reflecting the broader range of interactive components \system{} surfaces during browsing. No per-item difference reached statistical significance, indicating that \system{}'s additional in-situ reflection layer did not introduce meaningful usability costs relative to the baseline.

\subsection{Feed Alignment and Engagement}
\label{app:between-condition-methods}


To understand the alignment of each user's feed with their \emph{See More} and \emph{Avoid} preferences, as well as their engagement with this content, we need each video that appeared in the feed to be labeled as \emph{See More}, \emph{Avoid}, or neither. We could have used the labels provided by \system{} and the baseline, but chose not to because their classifiers differ slightly in order to accommodate their different feature sets. In particular, the \system{} classifier offers more granular classification, since users can attach examples to their preferences, and we leverage this granularity to aid decision-making. For this comparison to be fair, we want to use the same classifier, so we adopt the method from our technical evaluation that best represents both approaches: Name${\rightarrow}$LLM and Meta+Frames${\rightarrow}$LLM with zero examples provided.

We use the optimal threshold identified during our technical evaluation for this approach, $0.370$, which balances precision and recall. During the field study itself, however, both systems used higher, more precise thresholds (ranging from $0.4$ to $0.46$), because we experimented and found a preference for precision when the systems made decisions. Note that we would expect feed alignment and engagement for \emph{Avoid} content to be close to $0\%$, since both deployed systems filtered it out. The non-zero values we observe instead arise because our measurement classifier applies a less precise threshold.

\begin{table}[t]
  \centering
  \small
  \begin{tabular}{l c c c c c}
    \toprule
    Metric & \system{} & Baseline & $U$ & $p$ & $r$ \\
    \midrule
    \multicolumn{6}{l}{\emph{Feed alignment} (\% of watched videos matching $\geq$1 preference)} \\
    \quad \emph{See More} & $49.3 \pm 19.0$ & $20.3 \pm 14.9$ & $50.0$ & $\mathbf{.009}$ & $0.79$ \\
    \quad \emph{Avoid}    & $6.5 \pm 12.8$  & $19.0 \pm 14.9$ & $12.5$ & $.082$ & $-0.55$ \\
    \midrule
    \multicolumn{6}{l}{\emph{Engagement} (\% of total watchtime)} \\
    \quad \emph{See More} & $58.7 \pm 21.4$ & $21.0 \pm 11.6$ & $54.0$ & $\mathbf{.001}$ & $0.93$ \\
    \quad \emph{Avoid}    & $6.9 \pm 15.0$  & $16.7 \pm 12.4$ & $15.5$ & $.165$ & $-0.45$ \\
    \bottomrule
  \end{tabular}
  \caption{%
    {\normalfont Between-condition comparisons of feed alignment and engagement. Each watched video is labeled
    \emph{See More}, \emph{Avoid}, or neither; the \emph{See More} and \emph{Avoid}
    labels are independent, so a video may match both. \emph{Feed alignment} is
    video-level alignment; \emph{Engagement} is the share of each participant's
    total watchtime spent on videos of each category. \system{} and baseline columns report the mean\,$\pm$\,SD across participants.}
    \system{} feeds contained significantly more \emph{See More} content ($p{=}.009$) and \system{} users devoted a significantly larger share of watchtime to \emph{See More} content ($p{=}.001$). All other
    comparisons are non-significant.%
  }
  \label{tab:between-condition-levels}
\end{table}

\begin{table}[htbp]
  \centering
  \small
  \setlength{\tabcolsep}{5pt}
  \begin{tabular}{l c c c c}
    \toprule
    Metric & \system{} & Baseline & $\Delta$ (95\% CI) & $p_{\text{boot}}$ \\
    \midrule
    \multicolumn{5}{l}{\emph{Alignment slope} (\%-pts / session)} \\
    \quad \emph{See More} & $-0.13$ & $+6.13$ & $-6.26\ [-10.17,\,-2.17]$ & $\mathbf{.014}$ \\
    \quad \emph{Avoid}    & $+0.23$ & $-0.29$ & $+0.52\ [-5.18,\,+4.02]$ & $.698$ \\
    \midrule
    \multicolumn{5}{l}{\emph{Engagement slope} (\%-pts watchtime / session)} \\
    \quad \emph{See More} & $+0.90$ & $+6.19$ & $-5.30\ [-10.73,\,+0.38]$ & $.074$ \\
    \quad \emph{Avoid}    & $+0.73$ & $-0.99$ & $+1.72\ [-2.31,\,+4.09]$ & $.226$ \\
    \bottomrule
  \end{tabular}
  \caption{%
    {\normalfont Per-session trends in feed alignment and engagement, estimated
    with a linear mixed-effects model
    (\texttt{outcome $\sim$ session $\times$ condition + (session\,$|$\,participant)},
    fit by maximum likelihood; $n{=}15$ participants, 62 sessions). The slope
    columns are the model's per-condition fixed-effect slopes (\%-pts per session);
    $\Delta$ is the session\,$\times$\,condition interaction
    (\system{}\,$-$\,baseline), reported with a 95\% participant-level
    cluster-bootstrap confidence interval. $p_{\text{boot}}$ is a participant-level
    parametric-bootstrap likelihood-ratio test of the interaction
    ($\geq$2000 resamples of whole participants). The \emph{Avoid} engagement slope
    used a random-intercept structure (the random-slope fit did not converge); all
    other rows used the full random-slope model.}
    Baseline \emph{See More} alignment rose significantly faster than \system{}'s
    ($\Delta{=}-6.26$, $p_{\text{boot}}{=}.014$): \system{} feeds started aligned and stayed
    aligned, whereas baseline feeds drifted toward alignment over the study. All
    other trends are non-significant.%
  }
  \label{tab:between-condition-trends}
\end{table}

The between-condition comparisons in
Tables~\ref{tab:between-condition-levels} and~\ref{tab:between-condition-trends}
both rest on this single shared classification of each watched video. Within each
session we use the preferences a participant had active at that session's end and
compute watchtime as the sum of the $0.5$\,s heartbeats recorded in viewing events
on a video. A video is labeled \textit{See More} if it matches at least one active
\textit{See More} preference and \textit{Avoid} if it matches at least one active
\textit{Avoid} preference; the two labels are independent, so a video may carry
both or neither.

Table~\ref{tab:between-condition-levels} reports static levels, summarizing each
condition by the mean and standard deviation of a per-participant statistic.
\emph{Feed alignment} is the share of a participant's feed carrying each label:
pooling across all of a participant's sessions, we take the fraction of watched
videos carrying the \textit{See More} label and, separately, the fraction carrying
the \textit{Avoid} label. \emph{Engagement} is the share of watchtime each category
captured: for each participant and each category we sum watchtime over that
participant's videos in the category and divide by their total watchtime across all
watched videos, so a participant with no videos in a category contributes $0\%$ and
all 15 contribute. Each metric reduces to one value per participant, and the two
conditions ($n{=}8$ \system{}, $n{=}7$ baseline) are compared with a two-sided
Mann--Whitney $U$ test, with rank-biserial correlation as the effect size.
Table~\ref{tab:between-condition-trends} asks whether these measures trend over the study and whether the trend differs between conditions, using a linear mixed-effects model with session, condition, and their interaction. With baseline as the reference, the interaction $\Delta$ is the \system{}--baseline slope difference.


\begin{table}[htbp]
  \centering
  \small
  \begin{tabular}{l c c}
    \toprule
    Refinement sub-action & \system{} & Baseline \\
    \midrule
    Positive example added       & 53 (50 in-feed) & 0 \\
    Negative example added       & 23 & 0 \\
    Rubric bullet edited\textsuperscript{a} &  8 & 0 \\
    Preference renamed           &  2 & 0 \\
    Preference paused / resumed  &  6 & 1 \\
    \midrule
    Total              & 92 & 1 \\
    \quad\emph{per participant}  & 11.50 & 0.14 \\
    \bottomrule
  \end{tabular}
  \caption{%
    {\normalfont
    Preference-refinement actions by condition. \system{} refinement is dominated by example
    labeling (76 of 92, 83\%); only 10 of 92 (11\%) manually revise preference text.
    Note that adding an example automatically revised the preference's
    rubric (see the \emph{Preference description bullets} prompt in Appendix
    \ref{app:prompts}).
    The single baseline refinement was one pause/resume.
    Refinements were predominantly in-feed: the 50 in-feed positive examples plus
    all 23 negative examples sum to 73 of 92 (79\%) performed during browsing.%
    \\[2pt]
    \textsuperscript{a}\,Combines bullet add (2), edit (1), and delete (5).}%
  }
  \label{tab:refinement-breakdown}
\end{table}

\subsection{Preference Refinement}

To understand not just how often but in what ways participants refined their
preferences, we decompose the refinements into their constituent
actions (Table~\ref{tab:refinement-breakdown}). 
Participants could refine a preference in several ways: teach it by example (marking videos as positive or negative), edit the bullet-point rubric that defines it, rename it, or pause and resume it to refocus the feed. Marking a video as positive could be done within or outside the feed; marking one as negative could only be done within the feed; and all remaining actions could only be done manually, outside the feed.

\subsection{Preference Themes}

To characterize the kinds of content participants wanted to shape, we collected all 154 preferences created across both conditions (80 from \system{} participants and 74 from baseline participants), embedded them, and clustered the embeddings using $k$-means. We swept $k$ observing diminishing improvements in silhouette score beyond $k=15$; we selected $k=15$ as the value that produced clusters with clearly coherent themes upon qualitative inspection. After clustering, the first author manually inspected each cluster's contents and assigned the labels shown in Table~\ref{tab:clusters-k15}.

Several patterns emerge across these themes. Some clusters skewed sharply toward avoidance, particularly those targeting content participants found low-quality or aversive (\emph{Low-Quality Content}: 11 \emph{Avoid} vs. 2 \emph{See More}; \emph{Politics \& Conflict}: 7 vs. 2). Others skewed strongly toward \emph{See More}, concentrating around entertainment, lifestyle, and personal interests (\emph{Lifestyle Learning \& Personal Growth}: 22 vs. 5; \emph{Gaming}: 13 vs. 3; \emph{Football / Soccer}: 9 vs. 0). A few clusters were more balanced (e.g., \emph{Fashion \& Shopping}, \emph{Cars Racing \& Motorsports}, \emph{Specific YouTubers \& Creators}), reflecting that the same broad topic could appear as a \emph{See More} preference for one participant and an \emph{Avoid} for another.

\begin{figure*}[htbp]
    \centering
    \includegraphics[height=.75\textwidth, angle=90]{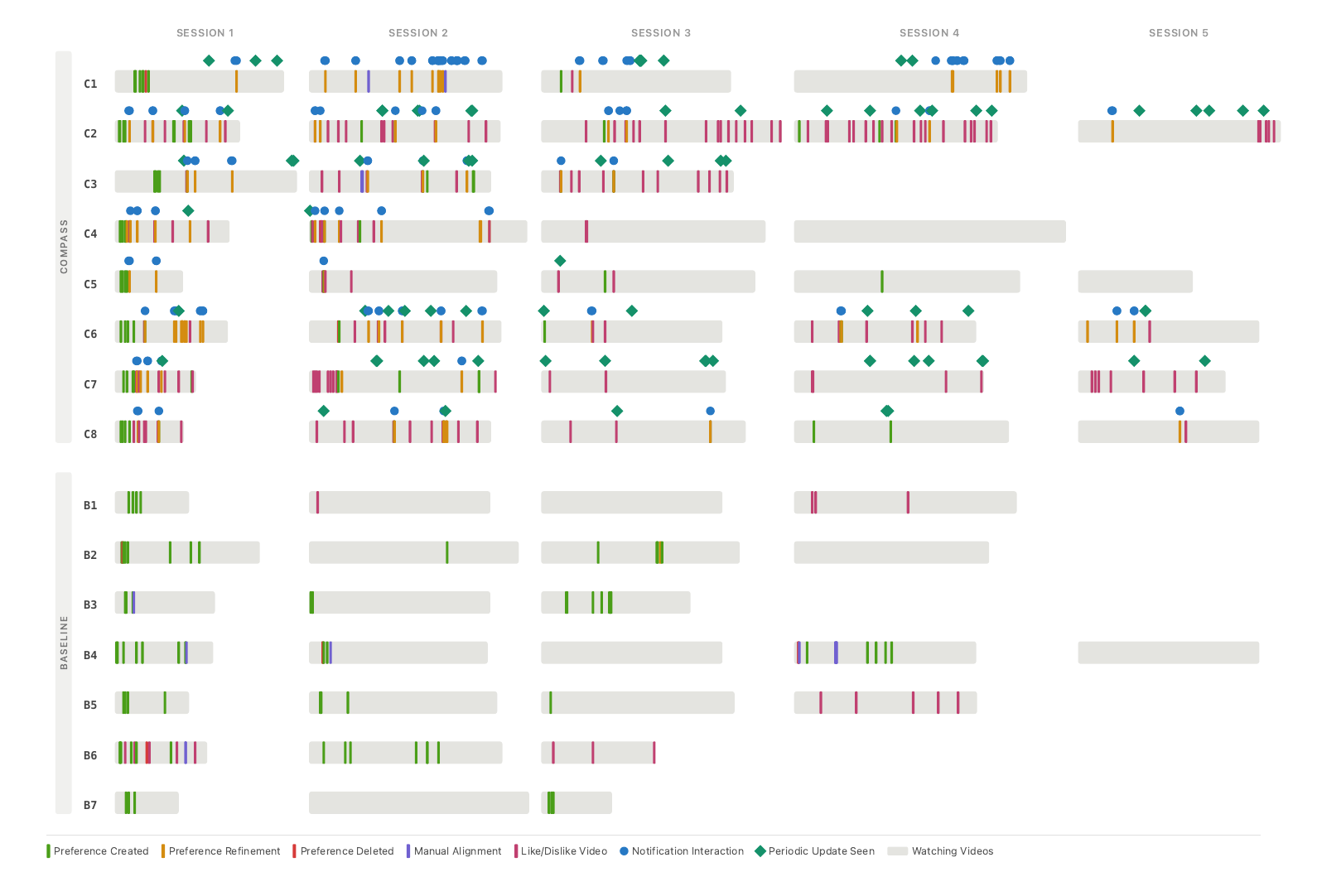}
    \caption{%
      \textbf{Session activity timelines for all \system{} (C1--C8) and Baseline (B1--B7) participants across up to five study sessions, showing the frequency and distribution of key interactions.}
      {\normalfont \emph{Events}: preference creation, refinement, and deletion; manual alignment; like or dislike video; notification interaction; and periodic update seen --- rendered as colored marks on or above the gray bar that represents time spent watching videos.}
      {\normalfont \emph{Notification interaction}: the user opening, confirming, or rejecting (i.e., providing negative feedback as opposed to simply dismissing) an in-situ reflection that appears when the current video matches a stored preference.}
  }
    \label{usage-timeline}
\end{figure*}

\begin{table*}[t]
\centering
\renewcommand{\arraystretch}{1.3}
\resizebox{\textwidth}{!}{%
\begin{tabular}{llllllllll}
\toprule
\textbf{ID} & \textbf{Condition} & \textbf{Sessions} & \textbf{Age} & \textbf{Gender} & \textbf{Occupation} & \textbf{Tech Proficiency} & \textbf{Platforms Used} & \textbf{Frequency} & \textbf{Daily Usage} \\
\midrule
C1 & Compass & 4 & 18--24 & Woman & Graduate Student & Intermediate & TikTok, Instagram Reels & Multiple times/day & 1--2 hours \\
C2 & Compass & 5 & 18--24 & Man & Student & Intermediate & TikTok, Instagram Reels, YouTube Shorts & Multiple times/day & 1--2 hours \\
C3 & Compass & 3 & 18--24 & Man & Student & Advanced & TikTok, Instagram Reels & Multiple times/day & 1--2 hours \\
C4 & Compass & 4 & 18--24 & Man & Financial Analyst & Intermediate & YouTube Shorts & Once a day & 30 min--1 hour \\
C5 & Compass & 5 & 18--24 & Woman & Medical Student & Advanced & YouTube Shorts & Once a day & {<}30 min \\
C6 & Compass & 5 & 18--24 & Man & Researcher & Expert & Instagram Reels & Multiple times/day & 30 min--1 hour \\
C7 & Compass & 5 & 18--24 & Woman & Medical Assistant & Intermediate & TikTok, Instagram Reels & Multiple times/day & {<}30 min \\
C8 & Compass & 5 & 25--34 & Man & Sales Consultant & Advanced & Instagram Reels, YouTube Shorts & Once a day & {<}30 min \\
\midrule
B1 & Baseline & 4 & 18--24 & Woman & Student & Advanced & TikTok, Instagram Reels, YouTube Shorts, Snapchat Spotlight & Multiple times/day & 4+ hours \\
B2 & Baseline & 4 & 18--24 & Man & Student & Expert & TikTok & Multiple times/day & 2--4 hours \\
B3 & Baseline & 3 & 18--24 & Woman & Social Media Coordinator & Advanced & TikTok, Instagram Reels, YouTube Shorts & Multiple times/day & 2--4 hours \\
B4 & Baseline & 5 & 18--24 & Man & Software Engineer & Expert & YouTube Shorts & Multiple times/day & 2--4 hours \\
B5 & Baseline & 4 & 18--24 & Woman & Unemployed & Intermediate & TikTok, Instagram Reels, YouTube Shorts & Multiple times/day & 2--4 hours \\
B6 & Baseline & 3 & 18--24 & Man & Student & Expert & Instagram Reels & Multiple times/day & 30 min--1 hour \\
B7 & Baseline & 3 & 18--24 & Man & Strength and Conditioning Specialist & Intermediate & TikTok, Instagram Reels & Multiple times/day & 30 min--1 hour \\
\bottomrule
\end{tabular}%
}
\caption{
\textbf{Complete participant demographics and social media usage across both study conditions.}
{\normalfont \system{} ($n=8$) used the full system with in-situ reflections and continuous feed alignment; Baseline ($n=7$) used a version without these ongoing supports, requiring participants to specify preferences and trigger feed alignment out of the feed. Sessions indicates the number of days each participant used the system.}
}
\label{tab:participants}
\end{table*}

\begin{table*}[t]
\centering
\footnotesize
\setlength{\tabcolsep}{10pt}
\begin{tabular}{r l p{7.5cm} r r}
\toprule
\textbf{ID} & \textbf{Cluster Label} & \textbf{Preferences} & \textbf{See More} & \textbf{Avoid} \\
\midrule
   0 & Politics \& Conflict & \textit{china, deadlock, guns, military videos, police/military, political podcasts, politics} & 2 & 7 \\
   1 & Lifestyle Learning \& Personal Growth & \textit{andrew huberman, animals, audio engineering, consumer psychology, cooking, dogs, educational content, electronic music production, entrepreneurial success stories, film, financial advice, freud, golf, hunting, korea travel, motivation, motivational speaking, music, outdoor adventure, outdoor adventure travel, podcasts, post grad life, psychology, see more preference, sexy girl, travel, women empowerment content} & 22 & 5 \\
   2 & Football / Soccer & \textit{arsenal, champions league, football commentary shows, football match highlights, football penalties, real madrid, rocket league, soccer, soccer history} & 9 & 0 \\
   3 & Fashion \& Shopping & \textit{clothes, clothing hauls, cota, fashion and wellness, men clothes, shopping} & 3 & 4 \\
   4 & Niche Interests \& Collectibles & \textit{film cameras, luxury watches, new york, old computers, religon, trivia, watchmaking} & 6 & 1 \\
   5 & Comedy \& Humor & \textit{comedy, funny, its always sunny in philadelphia, malcolm in the middle, stand up comedy} & 6 & 0 \\
   6 & Gaming & \textit{clash royale, competitive call of duty, deadlock game, gambling, gameing, gaming, geography games, home renovations, minecraft, minecraft speedrunning, programming, roguelike card games, slay the spire 2, tf2, the finals video game, video games} & 13 & 3 \\
   7 & Low-Quality Content & \textit{ai slop, booktok, brainrot, flame16anim, memes, movie clips, pranks, ragebait, sfm animation, sponsored content, subtitle sigma edit, wholesome edit} & 2 & 11 \\
   8 & Local City Content \& Travel & \textit{austin, austin texas, beatles history, new york travel, sxsw, taylor swift, things to do in la} & 6 & 2 \\
   9 & Cars Racing \& Motorsports & \textit{architecture, car, cars, food stores, formula 1, formula 1 crashes, formula one, guitar pedals, iceland, japan financial news, physics philosophy, racing, sfm} & 8 & 6 \\
  10 & Specific YouTubers \& Creators & \textit{@camman18, fadeon2k21, jschlatt, jynxzi, sidemen, skibidi toilet, swiftor says} & 3 & 4 \\
  11 & Fitness \& Wellness & \textit{exercise technique suggestions, gym advice, gym influencers, gym training exercises, sleep optimization, strength training, weightlifting techiniques, weightlifting techniques, workouts, yoga} & 9 & 1 \\
  12 & Funny \& Viral Clips & \textit{anime action series, car crash compilation, dog videos, funny dog, funny gaming videos, funny nintendo moments, minecraft parody comedy, rdcworld funny clips, roblox car crash compilation} & 6 & 3 \\
  13 & Food Recipes \& Coffee & \textit{healthy recipe, healthy recipes, lifestyle, new york coffee spots, new york restaurants, reciepes} & 6 & 0 \\
  14 & Art Design \& Math & \textit{drawing, geography, graphic design, math, sketchbook} & 5 & 1 \\
\bottomrule
\end{tabular}
\caption{%
    \textbf{15 preference themes identified via $k$-means clustering of participant-created preferences.}
    {\normalfont \emph{Labels}: clusters were manually labeled by the first author based on the preferences they contain.}
    {\normalfont \emph{Counts}: the \textbf{See More} and \textbf{Avoid} columns report the raw number of preferences in each cluster that participants marked as \textit{See More} and \textit{Avoid}, respectively.\strut}
}
\label{tab:clusters-k15}
\end{table*}

\subsection{Interaction Timelines}

Figure~\ref{usage-timeline} shows session activity timelines for all 15 participants across their study sessions. Each row corresponds to one participant, with up to five session columns. The gray bar represents time spent watching videos during that session, with width proportional to session length. Two types of marks are overlaid on each bar: vertical ticks on the bar denote events available in both conditions (preference creation, refinement, and deletion; manual alignment; like or dislike), while shapes above the bar denote events unique to \system{} (notification interactions and periodic update views).

Across the timelines, \system{} participants show visibly more preference refinement and like/dislike activity than Baseline participants, consistent with the significant differences reported in Section~\ref{sec:field-study-results}. The increase in preference refinement reflects \system{}'s in-situ interactions, which surface refinement opportunities directly in the feed rather than requiring participants to navigate to a separate preference page. The increase in like/dislike activity reflects \system{} repurposing these native controls as triggers for preference suggestion.

\begin{table}[htbp]
\centering
\footnotesize
\setlength{\tabcolsep}{2.5pt}
\begin{tabular}{lrrrrr}
\toprule
Type & Confirmed & Rejected & Swiped & Dismissed & Total \\
\midrule
Confident match (\textit{See More})  &  23 &   7 & 284 &  21 & 335 \\
Uncertain match (\textit{See More})  &  19 &  16 &  67 &   2 & 104 \\
Uncertain match (\textit{Avoid})     &   0 &   0 &   0 &   0 &   0 \\
Preference suggestion          &  18 & \na &  15 &  34 &  67 \\
\midrule
Total                          &  60 &  23 & 366 &  57 & 506 \\
\bottomrule
\end{tabular}
\caption{%
    \textbf{Interactive notification outcomes for \system{} participants (raw counts).}
    {\normalfont \emph{Notification types}: \emph{Confident match (See More)} is a low-prominence indicator shown alongside a video that \system{} matches to an existing \textit{See More} preference with high certainty, which the user can expand to confirm or correct. Confident matches for \textit{Avoid} preferences are filtered silently rather than surfaced, since surfacing them would defeat the purpose of avoidance. \emph{Uncertain match (See More)} and \emph{Uncertain match (Avoid)} are shown when a video falls near the similarity threshold for an existing preference, asking whether the user wants more content like this or wants it filtered out, respectively. \emph{Preference suggestion} is triggered when the user's in-the-moment behavior (e.g., disliking a video) suggests a new preference, proposing creation directly from the feed.}
    {\normalfont \emph{Outcomes}: \emph{Confirmed} indicates an explicit positive response. \emph{Rejected} indicates the user provided negative feedback that \system{} acts on, as opposed to \emph{Dismissed}, where the user simply closed the prompt without feedback. \emph{Swiped} means the user scrolled past the prompt without responding. Shaded cells (---) indicate structurally impossible outcomes where the UI does not offer that action for the given prompt type.\strut}
}
\label{tab:interactive-prompts}
\end{table}


\begin{table}[htbp]
\centering
\footnotesize
\setlength{\tabcolsep}{10pt}
\begin{tabular}{lccc}
\toprule
Outcome & Slope (\%-pts / session) & 95\% CI & $p_{\text{boot}}$ \\
\midrule
Confirmed & $+0.40$ & $[-4.78,\,+5.43]$ & $.883$ \\
\textbf{Rejected} & $\mathbf{-3.20}$ & $\mathbf{[-5.75,\,-0.99]}$ & $\mathbf{.001}$ \\
Swiped & $+3.83$ & $[-0.92,\,+8.49]$ & $.115$ \\
Dismissed & $-1.86$ & $[-4.90,\,+0.82]$ & $.219$ \\
\bottomrule
\end{tabular}
\caption{{\normalfont Temporal trends in interactive-notification outcome rates for
\system{} participants, estimated with a linear mixed-effects model
(\texttt{rate $\sim$ session + (session\,$|$\,participant)}, fit by maximum
likelihood; \system{} only, $n{=}8$ participants). The per-session rate is an
outcome's count divided by the interactive notifications displayed that session
(\%-points). The slope column is the model's fixed-effect session slope; the
$95\%$ CI and the two-sided $p_{\text{boot}}$ against zero come from a
participant-level cluster bootstrap that resamples whole participants with
replacement ($\geq$2000 times). Three of the four models retained a random slope;
the \emph{Rejected} model fell back to a random intercept because its random-slope
fit did not converge. The decline in rejection rate across sessions is
significant; no other outcome shows a significant trend.}}
\label{tab:notification-outcome-slopes}
\end{table}

\subsection{In-Feed Notifications}

\system{} inserts notifications into the feed to help users more casually reflect on their feed and its alignment with their stated preferences. Here we report quantitative metrics for how \system{} users responded to these notifications. This includes a general picture aggregated across the study, as well as temporal trends that reveal whether anything changed in how users responded over the course of 3--5 sessions.
Overall (Table~\ref{tab:interactive-prompts}), we find that the most common interaction was swiping (72.3\%). This was expected and supports our motivation that users want low-effort, casual interactions---including the ability to simply swipe past a notification when they do not find it relevant or feel compelled to respond. Even so, 16.4\% of responses provided feedback that \system{} could act on, suggesting that users did see opportunities where responding made sense.
We also find that users responded with actionable feedback far more often on uncertain matches than on confident ones (33.7\% vs.\ 9.0\%), meaning they felt more inclined to engage when \system{} expressed uncertainty than when it was certain---as we intended in our design.
There were also no uncertain matches for \textit{Avoid}. This is because users set fewer \textit{Avoid} preferences than \textit{See More}, and because our experimentally determined classification threshold was likely set too strictly. We discovered that users tended to write quite general \textit{Avoid} preferences (e.g., \emph{memes} or \emph{ai slop}).
For the temporal trends (Table~\ref{tab:notification-outcome-slopes}), we model each outcome's per-session rate with the same participant-level approach used for the feed-alignment and engagement slopes (a linear mixed-effects model with participant random effects, and a participant-level bootstrap for inference), here applied within the \system{} condition alone since the baseline issues no in-feed notifications. We find that all trends are non-significant except Rejected, which significantly decreased by $3.20$ percentage points per session ($\Delta = -3.20$, $95\%$ CI $[-5.75,\,-0.99]$, $p_{\text{boot}} = .001$). This suggests that the notifications increasingly matched what users wanted to act on.

\newtcolorbox{guidebox}[1][]{%
  enhanced, breakable,
  colback=gray!4,
  colframe=gray!55,
  boxrule=0.4pt,
  arc=1.5pt,
  left=5pt, right=5pt, top=4pt, bottom=4pt,
  IfValueTF={#1}{%
    title=#1,
    fonttitle=\bfseries\small,
    coltitle=black,
    colbacktitle=gray!14,
    titlerule=0pt,
    top=4pt,
  }{%
  },
}

\subsection{Study Interview Guides}
\label{app:study-guides}

All guides were semi-structured; indented items are follow-up probes
used at the interviewer's discretion.

\subsubsection{Pre-Study}
This guide was used before participants were introduced to their system.

\begin{guidebox}
\begin{enumerate}[leftmargin=1.6em, labelindent=0pt, labelsep=0.5em,
                  itemindent=0pt, topsep=0pt, partopsep=0pt]
  \item Can you describe how you typically use social media?

  \item Do you ever feel any frustration during or after using
        social media?
        \begin{enumerate}
          \item Do you feel frustrated because your feed does not
                align with your interests, or because it is so
                aligned that it makes you addicted?
        \end{enumerate}

  \item How often do you actively think about what kind of content
        you want to see in your feed?
        \begin{enumerate}
          \item When do you most often have these moments---while
                browsing your feed, or in retrospect?
        \end{enumerate}

  \item How well do you think social media understands what you
        want to see?
        \begin{enumerate}
          \item How do you think it learns your preferences?
        \end{enumerate}

  \item How much control do you feel over what you see in your
        social media feed?
        \begin{enumerate}
          \item Do you often leverage likes / dislikes / ``not
                interested'' to control your feed?
          \item Do you think these controls are effective?
        \end{enumerate}

  \item How much effort do you put into creating your ideal social
        media feed?
\end{enumerate}
\end{guidebox}

\subsubsection{Post-Study}
Separate versions were used for the baseline and \system{} conditions.
They share a common core, with additional items in the \system{} version
covering features unique to \textsc{Compass}.
However, participants in the baseline also thought they were using a system called \system{}.

\begin{guidebox}[Baseline Condition]
\begin{enumerate}[leftmargin=1.6em, labelindent=0pt, labelsep=0.5em,
                  itemindent=0pt, topsep=0pt, partopsep=0pt]
  \item Walk me through what it was like using \textsc{Compass}.

  \item Tell me about your process for creating preferences.
        \begin{enumerate}
          \item What motivated you to create preferences?
          \item Did you gradually change your preferences over time?
          \item Did you notice your feed changing based on the
                preferences you set?
        \end{enumerate}

  \item How was using \textsc{Compass} compared to your normal
        social media experience?

  \item Would you want to keep using a system like \textsc{Compass}?
        Why or why not?
\end{enumerate}
\end{guidebox}

\begin{guidebox}[\textsc{Compass} Condition]
\begin{enumerate}[leftmargin=1.6em, labelindent=0pt, labelsep=0.5em,
                  itemindent=0pt, topsep=0pt, partopsep=0pt]
  \item Walk me through what it was like using \textsc{Compass}.

  \item Tell me about your process for creating preferences.
        \begin{enumerate}
          \item What motivated you to create preferences?
          \item Did you gradually change your preferences over time?
          \item \textsc{Compass} includes a feature to add examples
                to preferences. Did you find that this helped teach
                \textsc{Compass} what your preference meant?
          \item Did you notice your feed changing based on the
                preferences you set?
        \end{enumerate}

  \item While you were scrolling, \textsc{Compass} showed you
        various notifications. Were there any specific notifications
        that stood out or that you found helpful?
        \begin{enumerate}
          \item Did you find the in-feed notifications helpful or
                intrusive?
        \end{enumerate}

  \item \textsc{Compass} periodically showed you updates for your
        preferences and your feed. Did you notice these updates, and did
        you find any of them helpful?
        \begin{enumerate}
          \item Did any of these updates help you reflect on your
                experience with YouTube Shorts?
        \end{enumerate}

  \item How was using \textsc{Compass} compared to your normal
        social media experience?

  \item Would you want to keep using a system like \textsc{Compass}?
        Why or why not?
\end{enumerate}
\end{guidebox}

\section{Technical Evaluation}

\begin{table*}[t]
\centering
\small
\setlength{\tabcolsep}{16pt}
\begin{tabular}{@{}llccccc@{}}
\toprule
\textbf{Preference Repr} & \textbf{Video Repr} & \textbf{ROC-AUC} & \textbf{Best F1} & \textbf{Precision} & \textbf{Recall} & \textbf{Thresh} \\
\midrule
Name                        & Meta                          & 0.857 & 0.719 & 0.800 & 0.652 & 0.292 \\
Name                        & Meta${\rightarrow}$LLM     & 0.886 & 0.720 & 0.759 & 0.685 & 0.289 \\
Name                        & Frames${\rightarrow}$LLM   & 0.881 & 0.713 & 0.772 & 0.663 & 0.302 \\
Name                        & Meta+Frames${\rightarrow}$LLM & 0.904 & 0.762 & 0.775 & \textbf{0.750} & 0.279 \\
Name${\rightarrow}$LLM   & Meta                          & 0.799 & 0.635 & 0.707 & 0.576 & 0.252 \\
Name${\rightarrow}$LLM   & Meta${\rightarrow}$LLM     & 0.831 & 0.644 & 0.842 & 0.522 & 0.393 \\
Name${\rightarrow}$LLM   & Frames${\rightarrow}$LLM   & 0.820 & 0.651 & 0.714 & 0.598 & 0.375 \\
\rowcolor{blue!8}
Name${\rightarrow}$LLM   & Meta+Frames${\rightarrow}$LLM & 0.842 & 0.675 & 0.794 & 0.587 & 0.370 \\
\midrule
\multicolumn{7}{@{}l@{}}{\textit{Include 1 video example in preference representation}} \\[2pt]
Name+Ex                     & Meta                          & 0.878 & 0.717 & 0.717 & 0.717 & 0.351 \\
Name+Ex                     & Meta${\rightarrow}$LLM     & 0.873 & 0.685 & 0.685 & 0.685 & 0.396 \\
Name+Ex                     & Frames${\rightarrow}$LLM   & 0.870 & 0.671 & 0.716 & 0.630 & 0.424 \\
Name+Ex                     & Meta+Frames${\rightarrow}$LLM & 0.904 & 0.731 & 0.771 & 0.696 & 0.410 \\
Name+Ex${\rightarrow}$LLM & Meta                          & 0.793 & 0.607 & 0.560 & 0.663 & 0.260 \\
Name+Ex${\rightarrow}$LLM & Meta${\rightarrow}$LLM     & 0.865 & 0.671 & 0.764 & 0.598 & 0.407 \\
Name+Ex${\rightarrow}$LLM & Frames${\rightarrow}$LLM   & 0.836 & 0.622 & 0.708 & 0.554 & 0.426 \\
\rowcolor{blue!8}
Name+Ex${\rightarrow}$LLM & Meta+Frames${\rightarrow}$LLM & 0.878 & 0.675 & 0.864 & 0.554 & 0.463 \\
\midrule
\multicolumn{7}{@{}l@{}}{\textit{Include 2 video examples in preference representation}} \\[2pt]
Name+Ex                     & Meta                          & 0.892 & 0.726 & 0.747 & 0.707 & 0.394 \\
Name+Ex                     & Meta${\rightarrow}$LLM     & 0.886 & 0.708 & 0.733 & 0.685 & 0.453 \\
Name+Ex                     & Frames${\rightarrow}$LLM   & 0.896 & 0.739 & 0.774 & 0.707 & 0.460 \\
Name+Ex                     & Meta+Frames${\rightarrow}$LLM & \textbf{0.913} & \textbf{0.792} & \textbf{0.940} & 0.685 & 0.493 \\
Name+Ex${\rightarrow}$LLM & Meta                          & 0.811 & 0.659 & 0.656 & 0.663 & 0.287 \\
Name+Ex${\rightarrow}$LLM & Meta${\rightarrow}$LLM     & 0.859 & 0.683 & 0.778 & 0.609 & 0.436 \\
Name+Ex${\rightarrow}$LLM & Frames${\rightarrow}$LLM   & 0.841 & 0.678 & 0.693 & 0.663 & 0.393 \\
\rowcolor{blue!8}
Name+Ex${\rightarrow}$LLM & Meta+Frames${\rightarrow}$LLM & 0.876 & 0.731 & 0.813 & 0.663 & 0.437 \\
\midrule
\multicolumn{7}{@{}l@{}}{\textit{LLM-as-Judge: Prompt LLM once per preference-video pair}} \\[2pt]
\multicolumn{2}{@{}l}{0 video examples} & {---} & 0.753 & 0.871 & 0.663 & {---} \\
\multicolumn{2}{@{}l}{1 video example}  & {---} & 0.748 & 0.859 & 0.663 & {---} \\
\multicolumn{2}{@{}l}{2 video examples} & {---} & 0.744 & 0.847 & 0.663 & {---} \\
\bottomrule
\end{tabular}
\vspace{2pt}
\caption{%
    \textbf{We compare different ways of representing preferences and videos before embedding them for similarity matching, evaluated on a test dataset of $N_{\text{eval}} = 408$ preference-video pairs.}
    {\normalfont \emph{Preference Repr}: the preference name alone, or expanded into a topic list via LLM (\textit{Name${\rightarrow}$LLM}), optionally augmented with different numbers of example videos.}
    {\normalfont \emph{Video Repr}: the video's metadata, keyframes, or both, optionally synthesized into a description via LLM (${\rightarrow}$LLM).}
    {\normalfont We also include a LLM-as-judge condition, where the LLM is prompted once per preference-video pair using all available information, representing a strong but computationally expensive upper bound.}
    {\normalfont We report ROC-AUC and the best F1 across cosine-similarity thresholds for each condition; \emph{Precision}, \emph{Recall}, and \emph{Thresh} give the precision, recall, and cosine-similarity threshold at that F1-maximizing operating point. \textbf{Bold}: best value per column (ROC-AUC, F1, precision, and recall) across all conditions. \colorbox{blue!8}{Blue shaded rows} indicate the configurations used by \system{}. We evaluate up to 2 examples, but \system{} enables up to 5 examples to be added per preference. For the per-pair LLM condition, predictions are binary without a continuous score, so ROC-AUC and a similarity threshold are not applicable (---).\strut}
}
\label{tab:ablations-full}
\end{table*}

\subsection{Dataset Construction}
\label{app:dataset-construction}

To construct the dataset used in our technical evaluation, we first embedded the preferences each participant wrote and the videos appearing in their feed (using a simple text representation of preference name and video metadata) and computed cosine similarity across all preference-video pairs. A naive random sample would be dominated by low-similarity videos, as most feed videos are irrelevant to any given preference. To address this, we adopted a stratified sampling strategy, binning videos by similarity score into four bins: $[0.0$--$0.1)$, $[0.1$--$0.2)$, $[0.2$--$0.3)$, and $[0.3+)$. Videos were sampled equally across bins to ensure moderate coverage of the similarity range. To keep the labeling task feasible, per participant, we automatically selected the combination of preferences and videos per bin that maximized total labeled samples while remaining within a cap of 50 labels. For example, a participant with 4 preferences and 4 relevance bins containing 3 videos each would label $4 \times 4 \times 3 = 48$ videos in total. A preference was only eligible if it had sufficient videos in every bin for the selected videos-per-bin count. For example, one participant's \textit{Avoid} preference for ``Freud'' was excluded as it had no videos in the $[0.3+)$ bin, consistent with the expectation that such content never appeared in their feed and would have produced no positive relevance labels. 

This sampling procedure yielded an original dataset of 708 labeled samples (185 positive, 523 negative) across 54 preferences. To evaluate example-augmented preference representations, we held out 2 positive examples per preference to be used, which required removing 11 preferences that had fewer than 2 positive labels. We further removed 7 preferences that had no remaining positives in the eval set after the holdout, as these could not contribute meaningful evaluation signal. The resulting evaluation dataset consists of 408 labeled samples (92 positive, 316 negative) across 36 preferences.

\subsection{Evaluation Results}
\label{detailed-results}
We present focused results in Table~\ref{tab:ablations-compact}. Complete results can be found in Table~\ref{tab:ablations-full}. \textbf{We found that the strongest video representations are produced by first using an LLM to synthesize metadata and keyframes before embedding.}
At $N=0$ examples, controlling for the preference representation consistently shows this advantage across conditions, as the LLM can leverage complementary textual and visual information to construct a more holistic representation for each video.

\textbf{Nevertheless, controlling for video representation and example count, directly embedding the preference name outperforms expanding it into a set of rubrics via an LLM before embedding.} This gaps is driven primarily by preferences with broader scope (e.g., \textit{lifestyle, cooking, comedy}) as opposed to more specific ones (e.g., \textit{Beatles history, Rocket League}). For broader preferences, LLM expansion casts too wide a net, matching videos within the topic but outside of the slice the user actually cares about (e.g., \textit{lifestyle} $\to$ \textit{healthy eating, physical fitness, sleep hygiene, stress management, work-life balance, \ldots}).
However, \system{} retains LLM expansion in its preference representation, as the generated subcategories serve a different purpose beyond matching: they give users a tangible vocabulary to inspect, edit, and refine their preferences

\textbf{In-feed annotations consistently improve preference matching, with gains becoming reliable at two positive examples.}
To assess this, we fix the best video representation from $N=0$ and vary the number of positive examples added to each preference. At $N=1$, gains are mixed: the no-LLM approach shows little change while the LLM-based preference approach sees modest improvement. At $N=2$, however, consistent gains emerge across both approaches, with the best configuration even surpassing LLM-as-Judge, a notable result given the cost and latency advantages of embedding-based approaches.

\lstdefinestyle{compasspromptlst}{%
  basicstyle=\footnotesize\ttfamily,
  breaklines=true,
  breakatwhitespace=false,
  breakindent=0pt,
  columns=fullflexible,
  keepspaces=true,
  showstringspaces=false,
  aboveskip=0pt,
  belowskip=0pt,
}

%
\newtcblisting{promptbox}[1][]{%
  enhanced,
  breakable,
  listing only,
  listing style=compasspromptlst,
  colback=gray!4,
  colframe=gray!55,
  boxrule=0.4pt,
  arc=1.5pt,
  left=5pt, right=5pt, top=4pt, bottom=4pt,
  IfValueTF={#1}{%
    title={#1},
    fonttitle=\footnotesize\itshape,
    coltitle=black!65,
    colbacktitle=gray!4,   
    titlerule=0pt,
  }{}%
}

\newcommand{\promptmodel}[1]{%
  \par\vspace{3pt}\noindent{\footnotesize\itshape\textcolor{black!65}{#1}}\par\vspace{1pt}}

\newcommand{\promptnote}[1]{%
  \par\vspace{1pt}\noindent{\footnotesize #1}\par\vspace{2pt}}

\section{Prompts}
\label{app:prompts}

This section lists every prompt sent to a large language model
in the \system{} system 
. Anything in curly braces (e.g.\ \texttt{\{preference\_name\}}) is a placeholder
that gets filled in when the prompt runs, not literal text. Embedding calls
(\texttt{text-embedding-3-large}) are omitted, as they carry no instruction
text. The title strip on each box names the model used for that call.



\paragraph{Preference-name suggestion.}
Converts a liked or disliked video into a 2--3 word interest topic.
\begin{promptbox}[Model: gpt-5.2]
A user {action_word} this video:

Channel: {channel}
Categories: {categories_str}
Tags: {tags_str}

Video Description:
{video_description}

Based on this video, suggest a slightly broader topic/interest category that the user might be interested in (if liked) or want to avoid (if disliked).

The topic should be:
- 2-3 words only
- Slightly broader than this specific video, but not too general
- Something that would match similar videos the user might also {action_word.replace('d', '')}

Respond with ONLY the topic name, nothing else.
\end{promptbox}
\promptnote{\textit{\{action\_word\}} is \texttt{liked} or \texttt{disliked}.}

\paragraph{Video description and keywords.}
Generates a video description plus matching keywords from metadata, with a
keyframe-aware variant when keyframe images are attached.
\begin{promptbox}[Model: gpt-4.1-nano \quad(with keyframe images)]
Based on the following information about a video and the provided keyframe images, generate:
1. A concise description of what the video is about
2. Exactly {settings.KEYWORD_COUNT} keywords for matching this video to user interests.

{metadata_text}
\end{promptbox}
\begin{promptbox}[Model: gpt-4.1-nano \quad(metadata only)]
Based on the following information about a video, generate:
1. A concise description of what the video is about
2. Exactly {settings.KEYWORD_COUNT} keywords for matching this video to user interests.

{metadata_text}
\end{promptbox}
\promptnote{\textit{\{metadata\_text\}} expands to a block of
\texttt{Title}, \texttt{Description}, \texttt{Channel}, \texttt{Categories},
and \texttt{Tags} lines.}

\paragraph{Preference description bullets.}
Authors or revises the bullet-point description of a preference. A first-time
variant seeds bullets; an update variant edits system-owned bullets while
preserving user-owned ones.
\begin{promptbox}[Model: gpt-5.2 \quad(new preference, no existing bullets)]
Preference title provided by user: {preference_name}

{examples_section}This is a new preference with no bullets yet. Generate bullet points describing what this preference is about.

Rules:
1. Generate up to 5 bullet points describing this preference
2. Keep bullets concise (10 words max)
3. Each bullet must represent a unique aspect of content
4. Use IDs starting from 1
5. Set owner to "SYSTEM" for all bullets

Return the bullet list as a JSON array with format: [{"id": 1, "owner": "SYSTEM", "text": "..."}]
\end{promptbox}
\begin{promptbox}[Model: gpt-5.2 \quad(update / regeneration)]
Preference title provided by user: {preference_name}

{examples_section}Update the bullet points to reflect the information the user has provided about their preference.

Rules:
1. Generate up to 5 bullet points describing this preference
2. Keep bullets concise (10 words max)
3. Each bullet must represent a unique aspect of content
4. You may ADD, EDIT, or DELETE bullets with owner "SYSTEM"
5. You may NOT modify or remove bullets with owner "USER"
6. For new bullets, use incrementing IDs starting from {next_id}

Current Bullets:
{current_bullets_json}

Return the updated bullet list as a JSON array.
\end{promptbox}
\promptnote{When example videos exist, \textit{\{examples\_section\}} is a
\texttt{Example videos of preference provided by user:} header followed by
\texttt{Video 1: \{desc\}}, \texttt{Video 2: \{desc\}}, \dots}

\paragraph{Preference keyword extraction.}
Extracts a fixed number of keywords for keyword-overlap matching against video
keywords.
\begin{promptbox}[Model: gpt-4.1-nano]
Preference: {preference_name}

{examples_section}Generate exactly {settings.KEYWORD_COUNT} keywords for matching videos to this user preference.
\end{promptbox}
\promptnote{Here \textit{\{examples\_section\}} is a list of
\texttt{Example video 1: \{desc\}}, \texttt{Example video 2: \{desc\}}, \dots}

\paragraph{Search-query generation.}
Produces diverse YouTube search queries from a context description, used to
fetch alignment content.
\begin{promptbox}[Model: gpt-5.2]
Generate {num_queries} diverse YouTube search queries based on this context:

Context: {search_context}

Requirements:
- Each query should search for similar content but use different keywords/angles
- Queries should be specific enough to return relevant results
- Queries should be diverse enough to return different video sets
- Each query should be 2-8 words
\end{promptbox}

\paragraph{Cluster labeling.}
Assigns a short thematic label to a cluster of video descriptions when
computing feed diversity.
\begin{promptbox}[Model: gpt-5.2]
Analyze these video descriptions and provide a single short thematic label (2-5 words) that captures what they have in common.

Video Descriptions:
{descriptions_text}

Requirements:
- Return ONLY the label, nothing else
- Keep it concise: 2-5 words
- Be specific but not overly narrow
- Use title case
- IMPORTANT: If the videos are clearly brand advertisements, corporate marketing, or have no meaningful content (e.g., ad campaign codes as titles, generic promo descriptions), label exactly as: "Ads/Promotional". Do NOT flag creator reviews, influencer content, or product showcases by individuals as ads.

Output format: A Single Label
\end{promptbox}
\promptnote{\textit{\{descriptions\_text\}} is up to five descriptions,
each truncated to 500 characters and numbered.}

\paragraph{Topic suggestion.}
Suggests new broad topics a user might enjoy, given their current interests,
avoid-list, and recent feed.
\begin{promptbox}[Model: gpt-5.2]
You are helping a user discover NEW content topics they might enjoy.

Here's what we know about this user:

Topics they are currently interested in:
{see_more_text}

Topics they want to avoid:
{avoid_text}

For additional context, here are the topics that have appeared in their recent feed:
{topics_text}

Based on this user profile, suggest {num_suggestions} NEW topics they might find interesting. These should be topics they haven't explicitly mentioned but would likely enjoy based on their interests.

Requirements:
- Return ONLY the topic suggestions, one per line
- Keep each suggestion concise (2-4 words)
- Suggest BROAD topic areas, not narrow subtopics or niche variations
- Suggest topics DIFFERENT from what's already in their feed or interests
- Do not suggest anything related to their avoid list
- Do not number the suggestions

Output format:
Topic 1
Topic 2
Topic 3
\end{promptbox}

\end{document}